\pdfoutput=1

\documentclass[twocolumn,letterpaper]{IEEEAerospaceCLS}  
\usepackage[utf8]{inputenc} 
\usepackage{textcomp} 

\usepackage[pdftex]{graphicx}    
\usepackage[colorlinks=true, urlcolor=blue, linkcolor=black, citecolor=black]{hyperref} 
\usepackage{enumitem}
\usepackage{multirow}
\usepackage{soul} 
\newcommand{\ignore}[1]{}  
\usepackage{tabularx} 
\usepackage{array} 

\begin{document}
\title{Integrated Co-design Framework for High-Efficiency Assembly and Integration of Small Satellites}

\author{%
Sen Hao Kam\\
National University of Singapore\\
5 Sports Drive, Singapore 117292\\
senhao90@nus.edu.sg
\and 
Kay Soon Low\\
National University of Singapore\\
5 Sports Drive, Singapore 117292\\
k.s.low@nus.edu.sg
\and
Sheral Crescent Tissera\\
National University of Singapore\\
5 Sports Drive, Singapore 117292\\
tsheral@nus.edu.sg
\thanks{\footnotesize 979-8-3195-1048-8/27/$\$31.00$ \copyright2027 IEEE}              
}

\maketitle

\thispagestyle{plain}
\pagestyle{plain}

\maketitle

\thispagestyle{plain}
\pagestyle{plain}

\begin{abstract}
\textbf{The rapid growth of small-satellite constellations has increased the need for higher manufacturing throughput, lower production costs, and shorter turnaround times. This makes efficient assembly, integration, and testing (AIT) increasingly important. Building on our prior work, this paper presents an integrated design-for-AIT approach for the 65 kg Lumelite-5 satellite. The integrated approach incorporates co-designing the satellite configuration, structural interfaces, subsystem placement, harness routing, and ground support equipment to facilitate efficient AIT. Service interfaces are centralized on one satellite side to simplify electrical ground support equipment (EGSE) access and routing. Skeletal support interfaces are developed to retain mounted components and service interfaces when removable enclosures are opened to expose internal harness-routing paths for inspection, troubleshooting, and rework. Defined mechanical references further support controlled removal and reinstallation of sensors and components. A removable double-hinge ground-support mechanism allows the top deck to open up to 180 degrees relative to the bottom deck while accommodating interconnecting-harness articulation, without introducing permanent hinge mass and complexity to the satellite. Finally, dedicated harness-routing and tie-down provisions provide predefined yet adjustable restraint locations to accommodate as-built routing variations and maintain controlled harness articulation during top-deck rotation. Collectively, these provisions improve physical accessibility, reduce unnecessary disturbance of integrated hardware, and provide a structured basis for repeatable AIT workflows through defined interfaces, controlled access, and structured harness routing. Through this work, we demonstrate how co-designing satellite architecture and AIT processes from the early development stages can support more scalable, repeatable, and efficient constellation production.}
\end{abstract} 

\tableofcontents

\section{Introduction}

\begin{figure}[!b]
\centering
\includegraphics[width=3.25in]{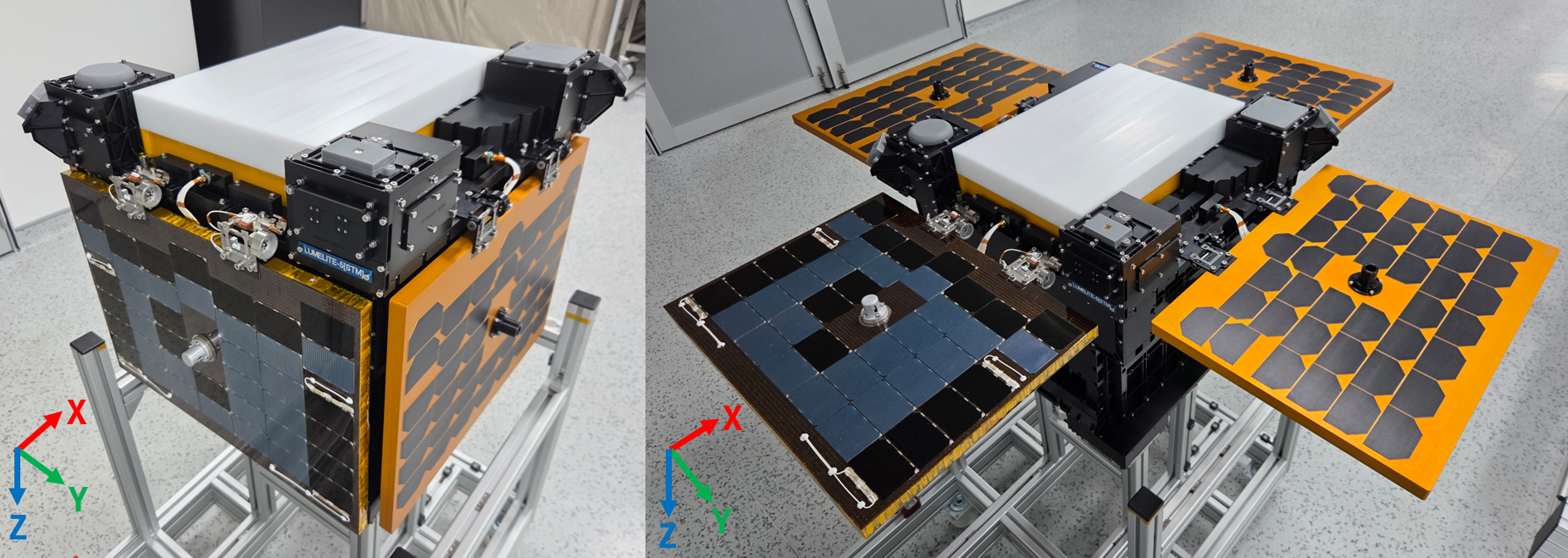}\\
\caption{Lumelite-5 structural model stowed (left) and fully deployed (right).}
\label{fig:fig1}
\end{figure}

Lumelite-5 is a 65~kg microsatellite (Fig.~\ref{fig:fig1}) developed by the Satellite Technology and Research Centre (STAR) in National University of Singapore (NUS). The mission of Lumelite-5 aims to demonstrate high-data-rate inter-satellite communication between a spacecraft in low Earth orbit (LEO) and a spacecraft in geosynchronous or medium Earth orbit (GEO/MEO). It was built based on the 12U CubeSats’s heritage of the Lumelite-1 to -4. Lumelite-4 was launched on 22 April 2023 into a 586~km near-equatorial orbit to demonstrate high-performance spaceborne VHF Data Exchange System (VDES) for maritime safety applications \cite{l4}. Lumelite-1 to -4 incorporated NUS’s patented modular and scalable satellite bus architecture \cite{patent}. Several of these qualified subsystems have been scaled and adapted for Lumelite-5. The Lumelite-5 bus incorporates a commercial-off-the-shelf (COTS) S-band transceiver as the primary communication system and an ultra-high-frequency (UHF) transceiver as the secondary communication system. The satellite is powered by four independently deployable solar panels, each retained by a hold-down and release mechanism (HDRM); and a lithium-ion battery pack (BP) with a nominal capacity of 397~Wh. The satellite’s attitude determination and control system (ADCS) comprises in-house-developed reaction wheels (RWs) and it’s electronics, together with COTS magnetorquers, a GPS receiver, coarse sun sensors (CSSs), fine sun sensors (FSSs), a magnetometer, and inertial measurement units (IMUs) \cite{adcs}. New developments for Lumelite-5 include a payload interface module (PIM) and an inter-satellite communication (ISC) payload with a software-defined radio (SDR), jointly developed by the Institute for Infocomm Research (I²R) from Agency for Science, Technology and Research (A*STAR), and Addvalue Technologies.

\begin{figure}
\centering
\includegraphics[width=3.25in]{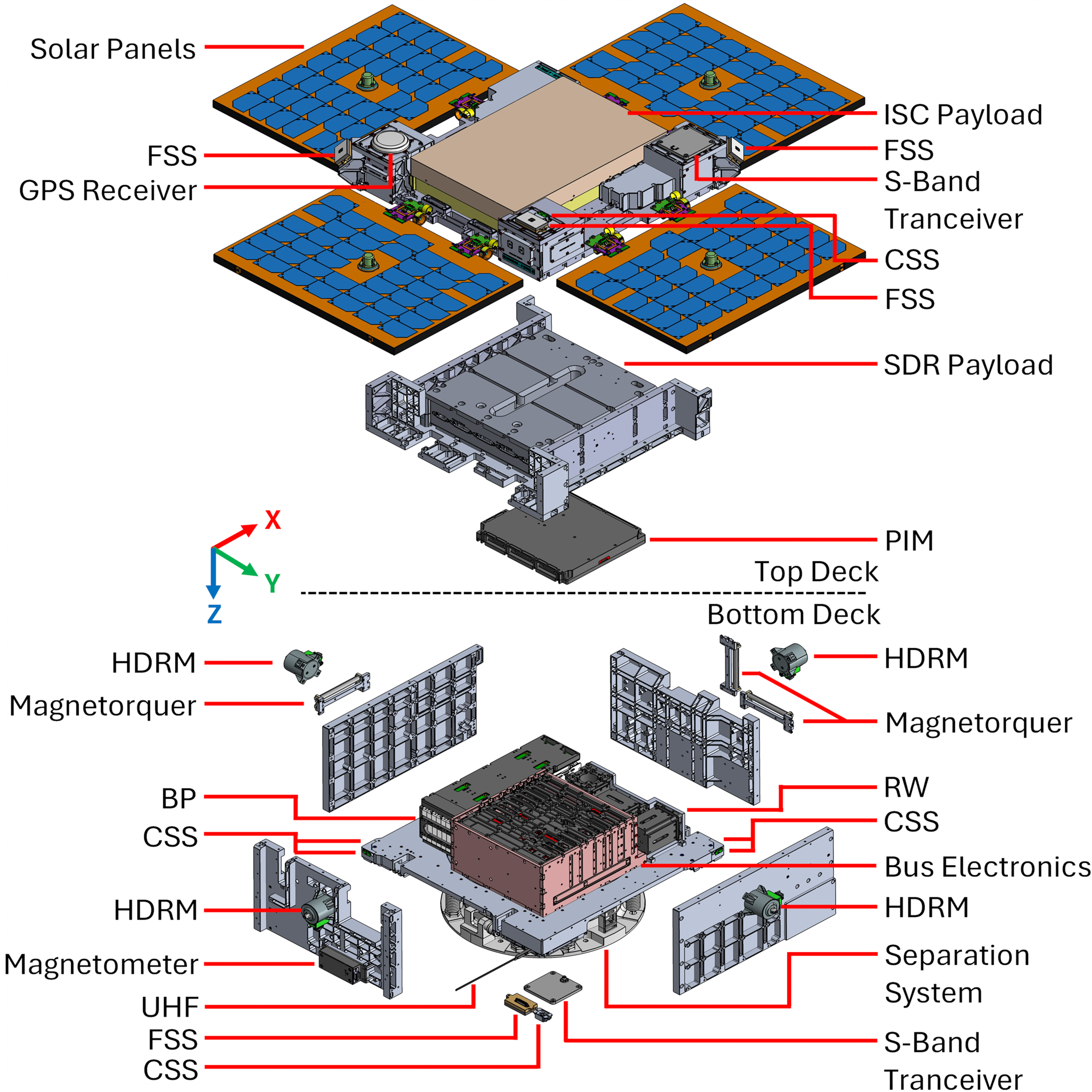}\\
\caption{Lumelite-5 system configuration.}
\label{fig:fig2}
\end{figure}
\begin{table}
\renewcommand{\arraystretch}{1.3}
\caption{\textbf{Lumelite-5 Specifications}}
\label{tab:tab1}
\centering
\begin{tabular}{|p{4cm}|p{3.6cm}|}
\hline
\bfseries Items & \bfseries Specifications \\
\hline\hline
Dimension, Stowed & 551~mm $\times$ 551~mm $\times$ 540~mm\\
Dimension, Deployed & 1184~mm $\times$ 1184~mm $\times$ 540~mm\\
Total Mass & 65~kg\\
Solar Power Output & 150~W peak\\
Battery Capacity & 397~Wh\\
Uplink/Downlink Data Rate & UHF up to 4800~bps \\
 & S-band up to 128 kbps \\
Pointing Accuracy & $<3.0^\circ$\\
\hline
\end{tabular}
\end{table}

During the preliminary design of Lumelite-5, the team conducted a literature review on the architectures and AIT experiences of previous small-satellite programs. Recurring challenges included restricted access to internal components and connectors, interference between harnesses and structural elements, panel-closeout difficulties, dependence on assembly sequence, and substantial disassembly when integrated hardware required troubleshooting or rework \cite{areda,kanji_norsat-1,hommy_nemo,gwozdecky_defiant}. Previous programs have also demonstrated the value of designing for maintainability and testability through improved access, modularity and standardized interfaces, built-in test capability, automation, and appropriately located test interfaces \cite{champlin_iridium,baghal_pnpsat-1}. These studies illustrate how satellite architecture can directly constrain physical access and the resulting AIT sequence.

These considerations are increasingly relevant to multi-satellite and constellation programs. Recent years have seen substantial activity in LEO constellation development and deployment \cite{koosha}. NASA’s 2026 State-of-the-Art of Small Spacecraft Technology report indicates that spacecraft platforms below 200 kg remain widely used while also noting increasing demand for larger SmallSat platforms \cite{nasa}. Kulu similarly reports growth in microsatellite and small-satellite constellation entries while identifying a broader shift toward larger satellite and, in some cases, smaller fleets \cite{kulu}. Regardless of the specific constellation architecture, recurring assembly and integration activities remain important when multiple satellites are produced. Design decisions that simplify access, assembly, rework, and test preparation therefore have the potential to reduce cumulative AIT effort and schedule impact.

In electronic systems, design-for-testability studies have emphasized considering modular product architecture and test-process development together from the early stages of product development \cite{kusiak}. A similar principle is reflected in concurrent-engineering approaches in the space sector, which emphasize the integrated development of space products and their associated life-cycle processes from the early design stage \cite{loureiro}. In addition, Costabile et al. highlighted the early involvement of test engineers in a concurrent-engineering environment to support reduced development iterations and shorter AIT schedules in satellite programs requiring high production rates \cite{costabile}.

Drawing on these lessons and our experience from Lumelite-4, this paper presents an integrated co-design approach for AIT. In this paper, we use the Lumelite-5 structural model (SM) as a representative example to establish and refine the satellite assembly and integration procedures, including subsystem accessibility, mechanical interfaces, and harness routing. The satellite architecture, subsystem placement, structural interfaces, harness routing, and ground-support equipment were developed concurrently instead of independent activities. The present work focuses primarily on assembly and integration; testability provisions are discussed in terms of their intended functions and implementation value because the overall Lumelite-5 test campaign has not yet been completed.

The main contributions of this paper are as follows:
\raggedbottom
\begin{enumerate}[label=\arabic*., leftmargin=*, align=left]
\item \textbf{Integrated co-design framework:} Four coordinated techniques---centralized service interfaces, skeletal support interfaces, a removable double-hinge MGSE, integrated and modular tie-down retainers---are developed together with the satellite architecture, subsystem placement, harness routing, and ground-support equipment rather than as separate AIT provisions (Sections~2 and~3).
\item \textbf{Hardware-level implementation on a microsatellite SM:} The framework is implemented and evaluated through integration activities on the 65~kg Lumelite-5 SM, including centralized interface accessibility, top-deck opening to 90$^\circ$ and 180$^\circ$, retention of most bus-level functionality in the open configuration, and assessment of interconnecting-harness articulation against the applicable RF-cable dynamic bend-radius requirement (Sections~2 and~3).

\item \textbf{AIT implementation value, limitations, and future extensions:} The individual and combined techniques are assessed for progressive assembly, functional verification, troubleshooting, rework, and final satellite closeout. Remaining design limitations are identified, a conceptual satellite and harness reconfiguration is investigated, and a pathway toward future assisted and automated AIT is developed (Sections~3--5)
\end{enumerate}
Through these contributions, this paper shows how satellite architecture and AIT processes can be co-developed from the early design stage to preserve accessibility, control harness routing, and support more repeatable integration workflows while providing a basis for future AIT automation. The remainder of this paper is organized as follows. Section~2 describes the integrated co-design framework implemented on Lumelite-5, while Section~3 discusses the individual and combined AIT benefits and implementation value. Section~4 and~5 presents the remaining limitations, the conceptual improvement study, and the pathway toward future AIT automation. Section~6 concludes the paper.

\section{Integrated Co-Design Framework}
\begin{figure}
\centering
\includegraphics[width=3.25in]{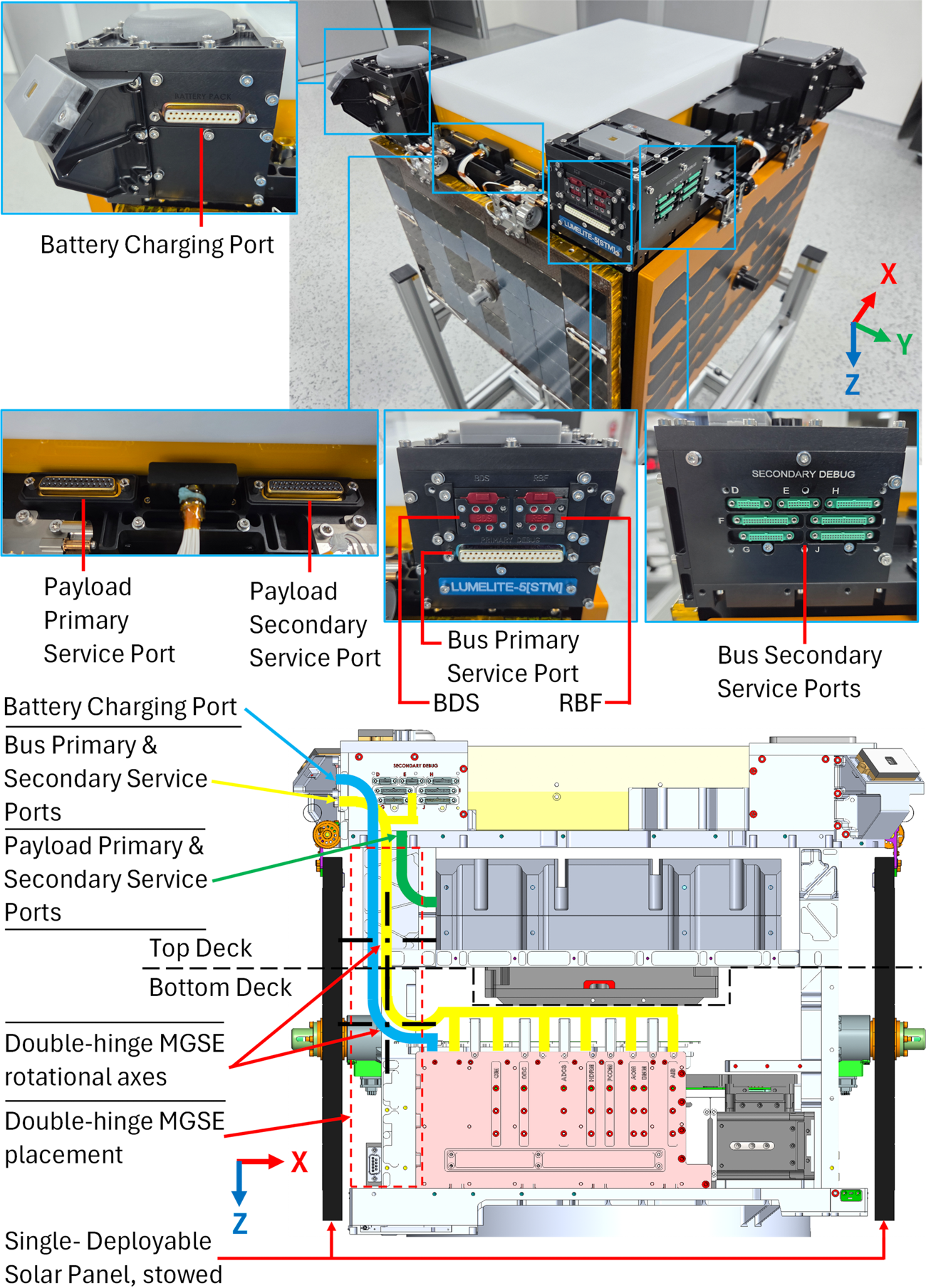}\\
\caption{Lumelite-5 $-X$ side showing the centralized service ports, RBF interface, BDS, battery charging port, and a simplified representation of the corresponding harness routing.}
\label{fig:fig3}
\end{figure}
In this section, we present four techniques within the proposed co-designed framework that were tested on the Lumelite-5 SM, which served as the development platform for refining the satellite assembly and integration process. During early integration development, low-cost 3D-printed representations reproduce the mechanical envelopes and mounting interfaces of selected sensors and components, allowing the assembly sequence, subsystem accessibility, mechanical interfaces, and harness routing to be evaluated and iterated without unnecessary handling of flight hardware. Once the assembly procedure and harness routing have been sufficiently refined, these representations are replaced by representative dummy units for subsequent AIT activities. The four co-designed techniques are detailed below.
\subsection{Technique 1: Centralized Service Interfaces}
Accessibility during thermal-vacuum testing and launch-vehicle upper-stage integration drove the requirement to centralize the bus and payload service interfaces together with the battery charging interface. The specific location of these centralized interfaces on the satellite $-X$ side was then determined by the placement of the bus electronics, the associated harness-routing architecture, and the double-hinge MGSE geometry. Considering these constraints together, the upper region of the $-X$ side was selected as the most suitable location for maintaining interface accessibility while supporting practical harness routing and top-deck rotation. In this paper, service interfaces refer to the EGSE-accessible ports used for functional checkout, troubleshooting, and satellite-health monitoring, while the separate battery charging interface is dedicated to charging the battery while the satellite remains fully assembled and stowed. As shown in Figure~\ref{fig:fig3}, the primary, secondary, and payload service ports, remove-before-flight (RBF) interface, battery disconnect switch (BDS), and battery charging port are centralized on the satellite $-X$ side. A simplified representation of the corresponding harness routing is also included in the satellite side view to illustrate the routing concept.

\begin{figure}
\centering
\includegraphics[width=3.25in]{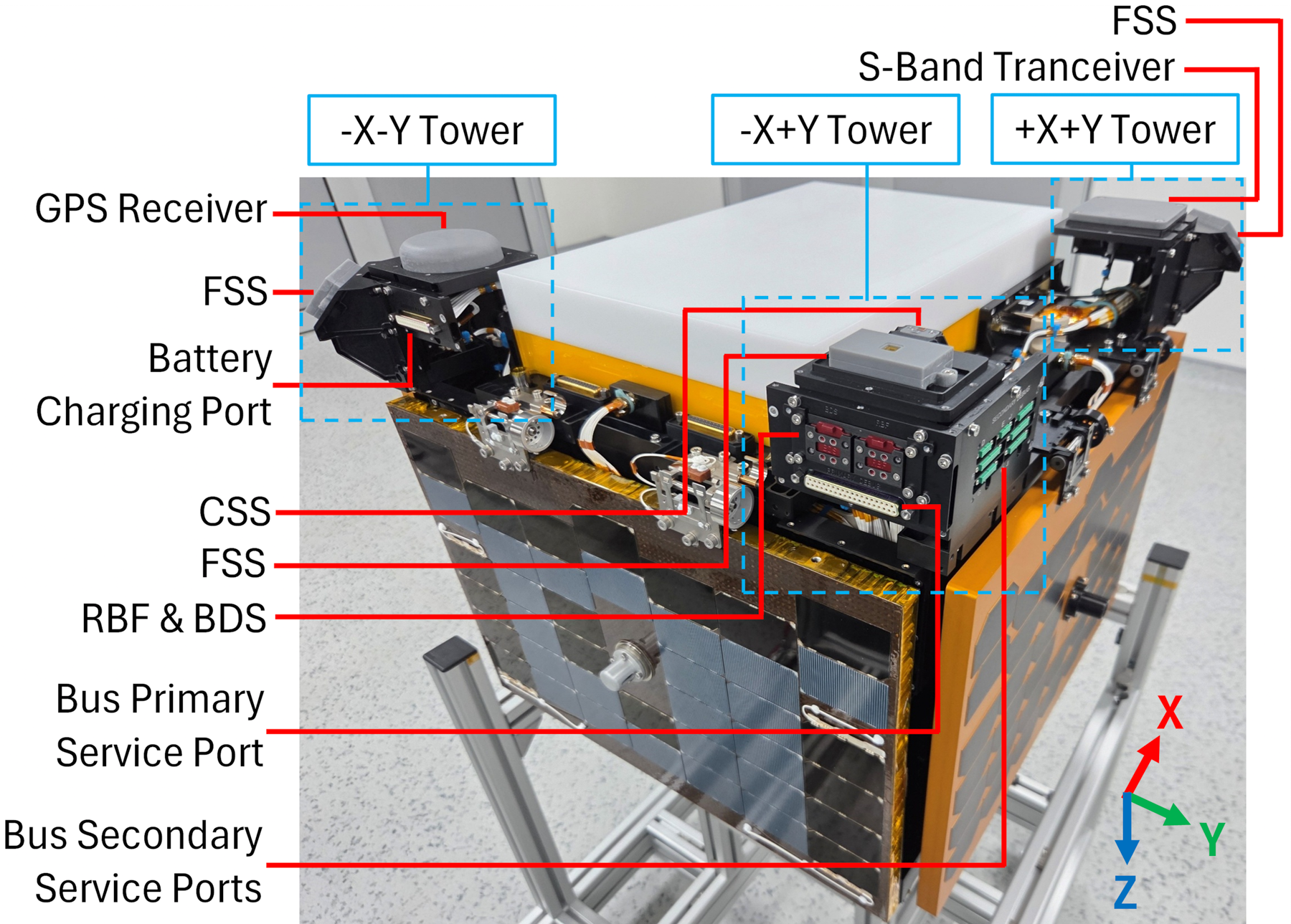}\\
\caption{Skeletal support interfaces implemented at the $-X-Y$, $-X+Y$, and $+X+Y$ towers with the enclosures removed, showing the components and service interfaces retained on the skeletal structures and the exposed internal harness-routing paths.}
\label{fig:fig4}
\end{figure}

The interface locations were co-developed with the harness-routing architecture and double-hinge MGSE geometry. Harnesses connecting the top and bottom decks are routed close to the rotational axes of the double-hinge mechanism. During top-deck rotation, locating the harnesses near these axes minimizes the change in harness path length and consequently the additional harness length required to accommodate the relative deck motion. The design and operation of the double-hinge mechanism are described in the Double-Hinge MGSE subsection.

\begin{figure}
\centering
\includegraphics[width=3.25in]{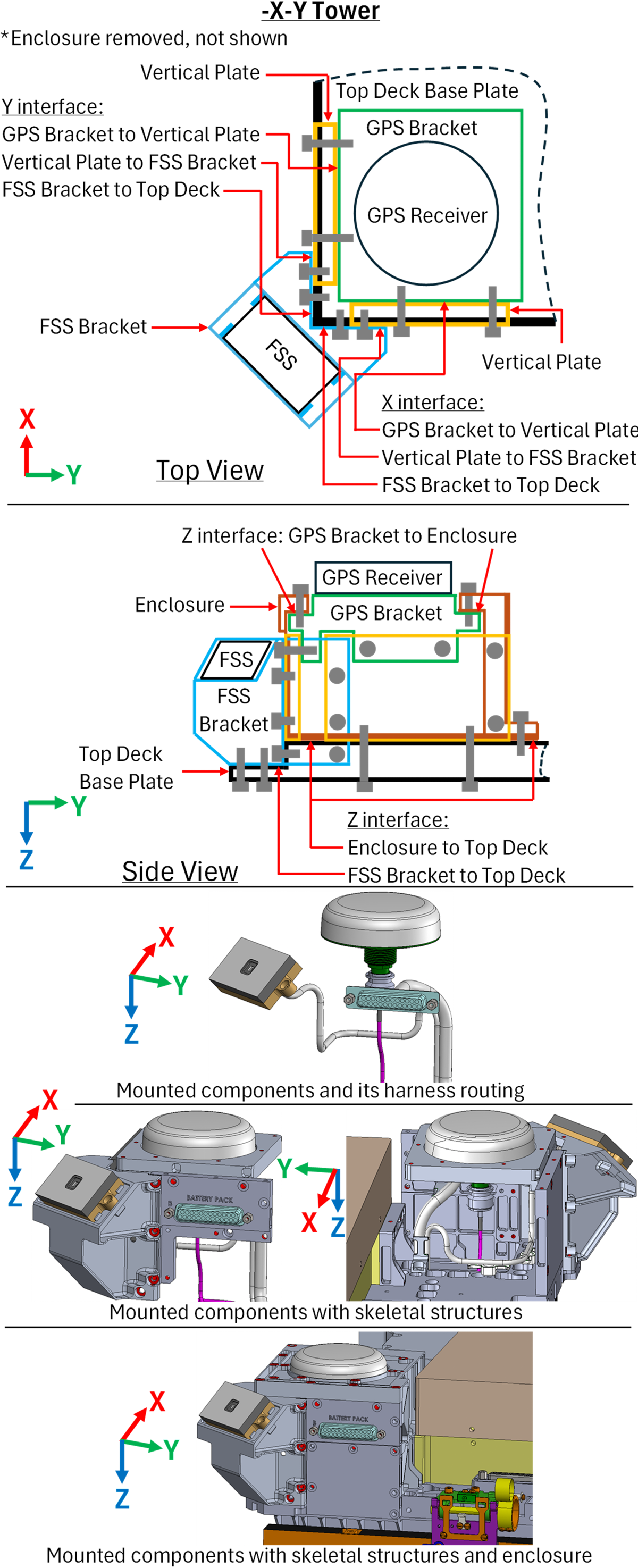}\\
\caption{Simplified top and side views of the $-X-Y$ tower mechanical interfaces, with corresponding CAD views of the component-bracket arrangement and integrated tower configuration.}
\label{fig:fig5}
\end{figure}
\subsection {Technique 2: Skeletal Support Interfaces}

Previous spacecraft programs provide examples of integration-access constraints arising from component placement and structural architecture. NORSAT-1 required a specific panel assembly sequence to preserve access as integration progressed \cite{kanji_norsat-1}, while in NEMO, relocation of the $-Y$ sun sensor from the $-Y$ panel to the SDR enclosure reduced the hardware mounted on that panel and simplified its assembly \cite{hommy_nemo}. The NEMO modification involved the relocation of a single component; however, accommodating a larger number of components and interfaces within a similarly constrained region would introduce additional packaging and integration complexity.

For Lumelite-5, much of the available top-deck area is occupied by the ISC payload as shown in Figure~\ref{fig:fig4}, payload harness routing, payload service and solar-panel interfaces. The remaining space must accommodate the sun sensors, GPS receiver, S-band transceiver, bus service and battery charging interfaces. Considering the field-of-view requirements of the sun sensors, the antenna-visibility requirements of the GPS receiver and S-band transceiver, and the limited top-deck area, a tower-based arrangement was adopted at three corners of the top deck, as shown in Figure~\ref{fig:fig4}. The CSS, FSS, GPS receiver, and S-band transceiver are positioned toward the upper portions of the towers, while the bus service and battery charging interfaces are accommodated on the side-facing regions.

To retain access for component installation and harness routing within these constrained regions, the team developed a new concept, referred to here as skeletal support interfaces, for the $-X-Y$, $-X+Y$, and $+X+Y$ towers. Each tower comprises skeletal support structures and a removable enclosure. The service interfaces and externally mounted components remain mounted on the skeletal structures rather than on the removable enclosures, allowing the enclosures to be removed without removing the associated components and interfaces. Beyond preserving integration access, the skeletal-support arrangement also defines the mechanical references used to locate the externally mounted components.

\begin{figure}
\centering
\includegraphics[width=3.25in]{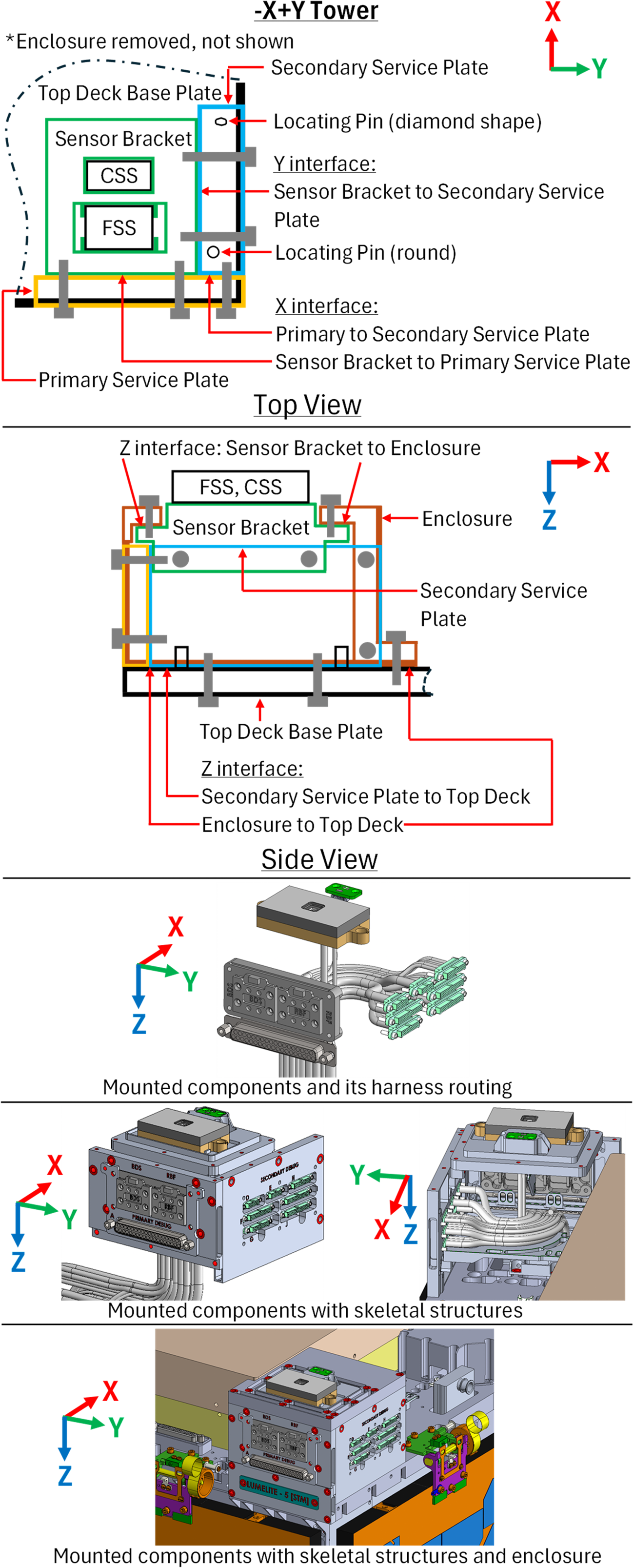}\\
\caption{Simplified top and side views of the $-X+Y$ tower mechanical interfaces, with corresponding CAD views of the sensor- and service-interface arrangement and integrated tower configuration.}
\label{fig:fig6}
\end{figure}

Figure~\ref{fig:fig5} shows simplified top and side views of the $-X-Y$ tower mechanical interfaces. The FSS bracket is referenced directly to the top-deck base plate in the X, Y, and Z directions, establishing its position and orientation relative to the satellite structure. The GPS bracket is subsequently referenced to the FSS bracket in the X and Y directions through the vertical plates interfaces. In the Z direction, the GPS bracket is referenced to the enclosure, which establishes its vertical position. A similar interface arrangement is implemented at the $+X+Y$ tower. A different reference arrangement is used for the $-X+Y$ tower, as shown in Figure~\ref{fig:fig6}. The secondary service plate is located relative to the top-deck structure using a round-and-diamond locating-pin pair, which establishes its position in the X-Y plane. The sensor bracket carrying the FSS and CSS is then positioned relative to the secondary service plate. The secondary service plate establishes the sensor-bracket position in the Y direction, while the primary service plate provides an assembly reference surface for establishing the X-direction position. In the Z direction, the sensor bracket is referenced to the enclosure, which establishes its vertical position. These interfaces define the component positions while allowing the enclosure to remain a removable structural element.
\subsection {Technique 3: Double-Hinge MGSE}
Hinged-panel architecture has previously been used to preserve internal access during spacecraft integration. PnPSat-1 employed six hinged modular panels that provided access for installation of internal components before spacecraft closeout \cite{baghal_pnpsat-1,pnp}. The IXO spacecraft-bus concept similarly incorporated equipment panels that could fold down or hinge open during integration and test to provide access to internal avionics \cite{xray}.

A single-hinge arrangement can provide effective integration access when sufficient packaging space and clearance are available to position the hinge axis and route the interconnecting harnesses with adequate bend radius. However, in a more compact spacecraft, limited available space can constrain the hinge location and accessible opening geometry. Harnesses routed across a single hinge axis must also accommodate the full angular displacement between the two structural sections, which may result in tighter bending around the hinge or require additional packaging space to accommodate the required harness routing length.

\begin{figure}[!t]
\centering
\includegraphics[width=3.25in]{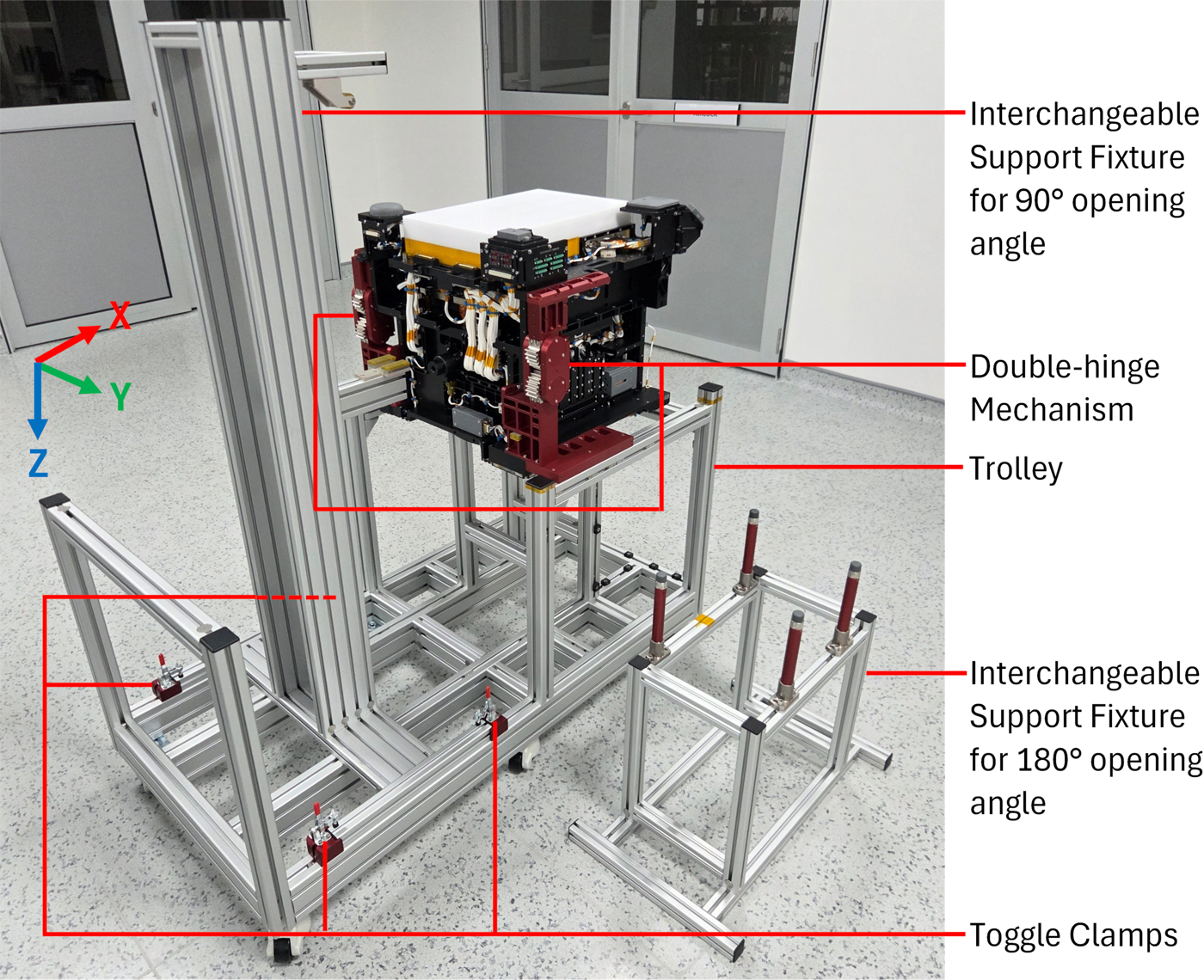}\\
\caption{Lumelite-5 double-hinge MGSE and interchangeable top-deck support fixtures.}
\label{fig:fig7}
\end{figure}

These considerations motivated the development of a double-hinge mechanical ground support equipment (MGSE) mechanism for Lumelite-5, using two rotational axes to allow the top deck to be opened relative to the bottom deck. Figure~\ref{fig:fig7} shows the Lumelite-5 integration MGSE. The satellite is mounted to the trolley through its separation-system interface. Prior to installation of the double-hinge mechanism, the four solar panels are disconnected from their respective bulkhead connectors and removed from top deck. Removal of the solar panels provides the clearance required for installation of the double-hinge mechanism and provides unobstructed hinge rotation. Interchangeable support fixtures mounted to the trolley using four toggle clamps hold the top deck at opening angles of 90$^\circ$ and 180$^\circ$.

The double-hinge mechanism incorporates a pair of 1:1 spur gears to synchronize the two hinge axes. As the top deck is opened, its angular displacement is distributed across both hinge axes, allowing controlled rotation to an opening angle of 180$^\circ$. The interconnecting harnesses consequently articulate across two rotational axes rather than undergoing the entire angular change around a single axis. The mechanism is implemented as MGSE rather than as part of the flight structure and is therefore used only during ground assembly and integration.

\begin{figure}
\centering
\includegraphics[width=3.25in]{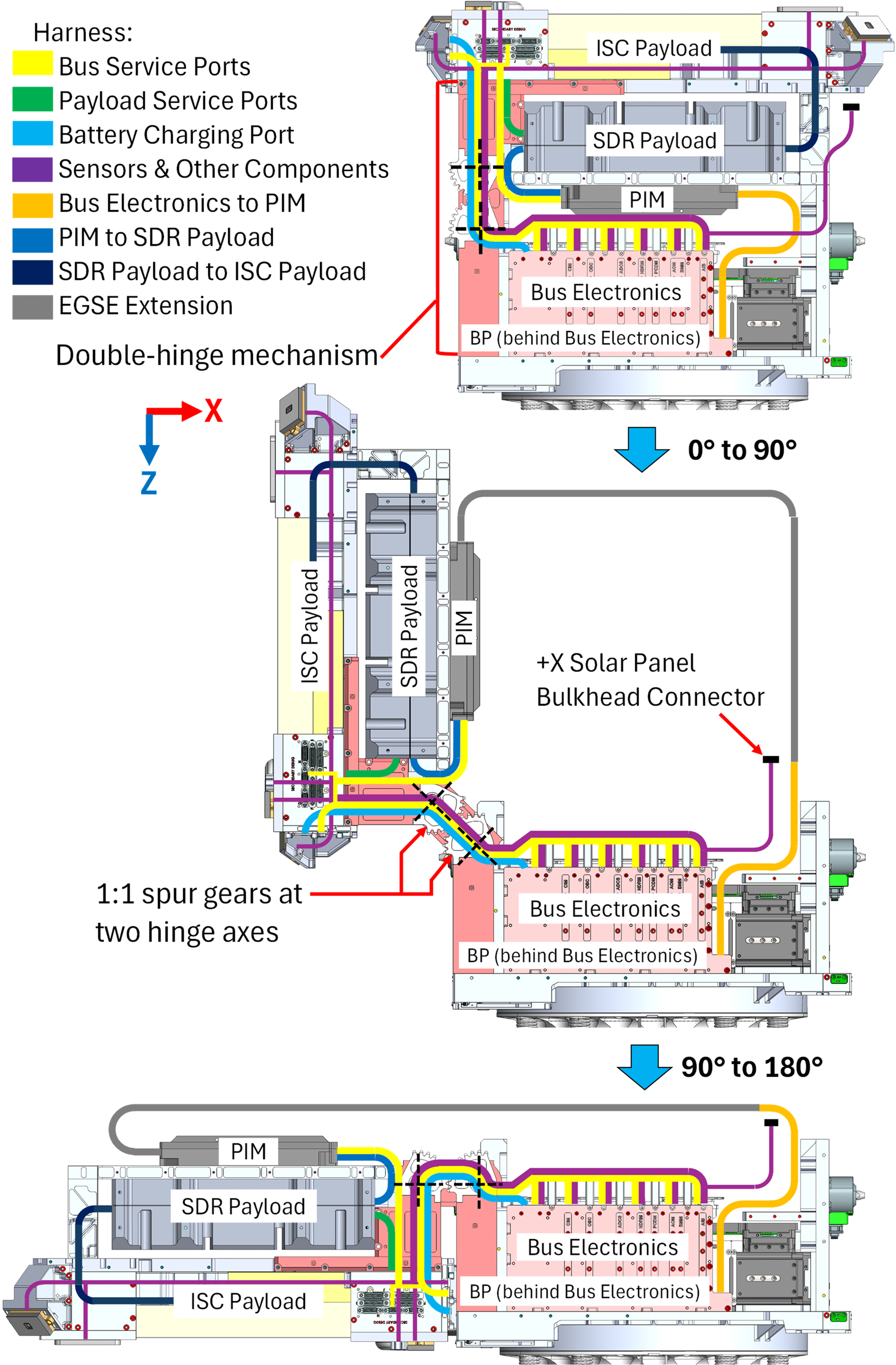}\\
\caption{Simplified side views of top-deck rotation from 0$^\circ$ to 180$^\circ$, showing interconnecting-harness articulation and EGSE extension routing.}
\label{fig:fig8}
\end{figure}

\begin{figure*}
\centering
\includegraphics[width=6.5in]{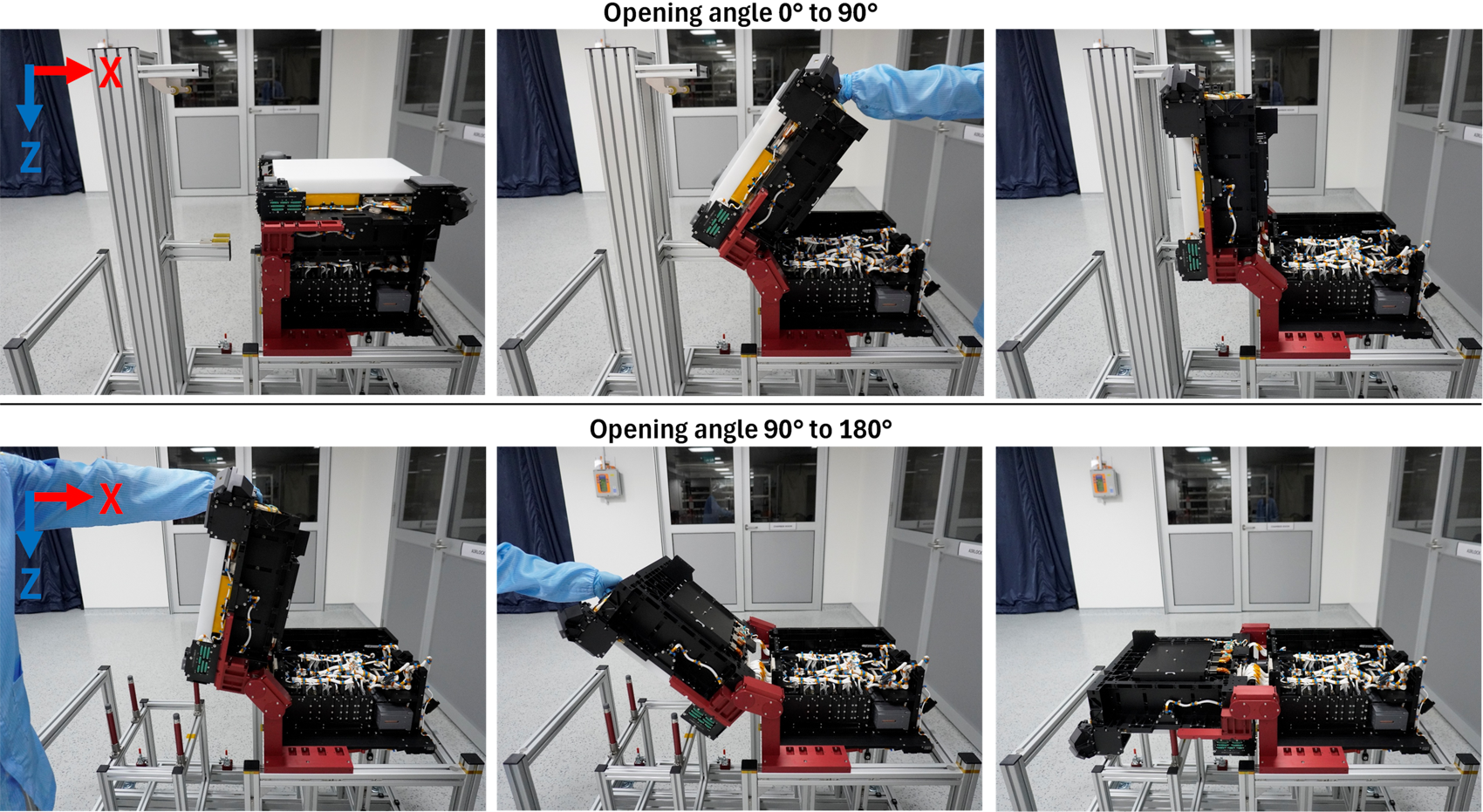}
\caption{Sequence of top-deck rotation from 0$^\circ$ to 180$^\circ$ using the double-hinge MGSE.}
\label{fig:fig9}
\end{figure*}

Before top-deck rotation, the $+X$ solar-panel bulkhead connector is dismounted from the top-deck structure, and the bus-electronics-to-PIM harness is disconnected from the PIM. Figure~\ref{fig:fig8} illustrates the interconnecting harness configuration during rotation from 0$^\circ$ through 90$^\circ$ to 180$^\circ$. Harnesses that remain connected between the two decks articulate around the two hinge axes while EGSE extension harness are being used to reconnect the bus electronics to the PIM. Although the solar panels remain disconnected, the majority of the satellite subsystems remain electrically connected and available for functional activities while the top deck is open. Figure~\ref{fig:fig9} shows the corresponding physical top-deck rotation sequence.
\subsection {Technique 4: Integrated and Modular Tie-Down Retainers}
\begin{figure}
\centering
\includegraphics[width=3.25in]{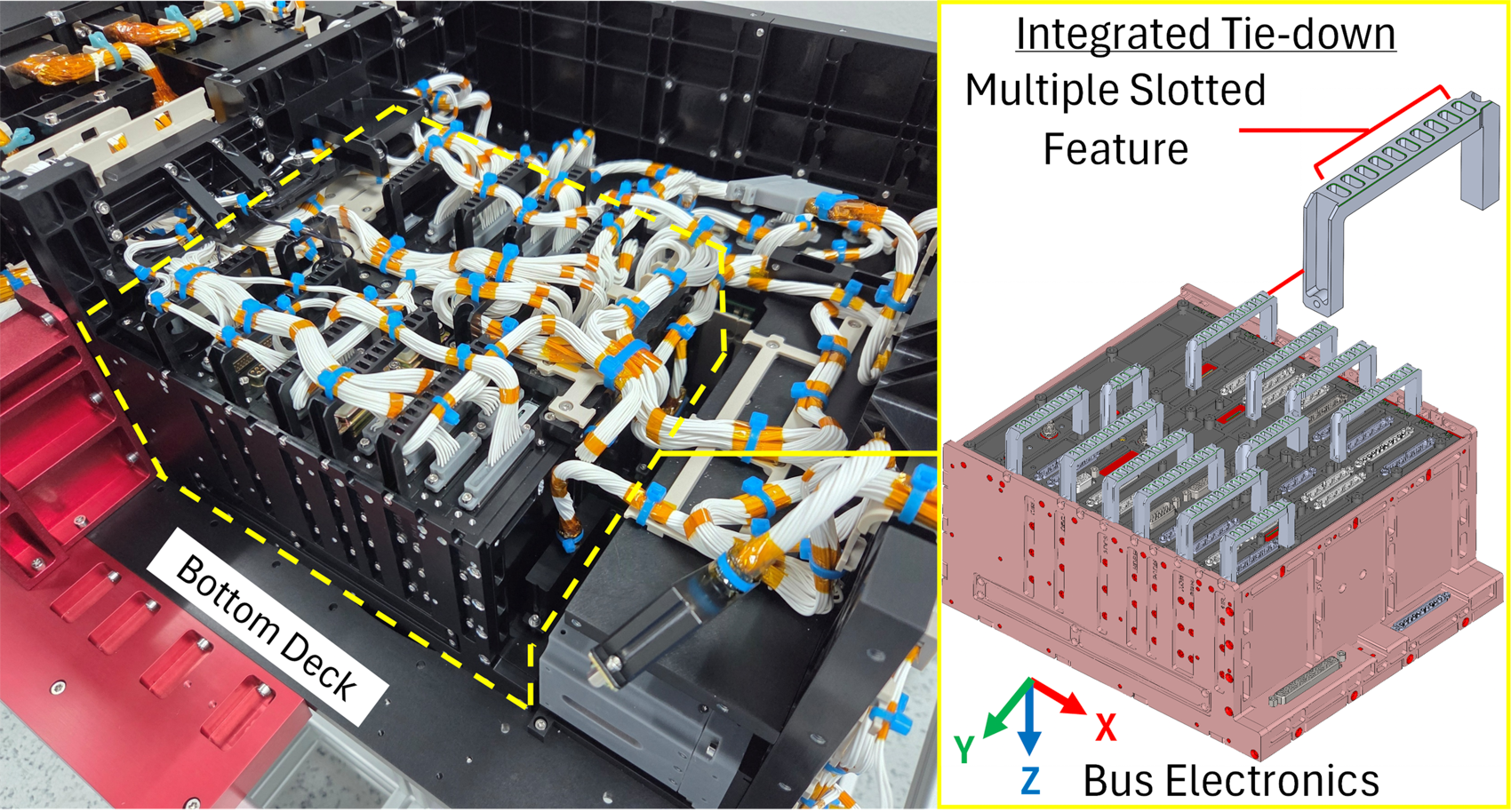}\\
\caption{Lumelite-5 bottom deck highlighting the bus electronics and associated aluminum-alloy integrated harness tie-down brackets.}
\label{fig:fig10}
\end{figure}
\begin{figure}
\centering
\includegraphics[width=3.25in]{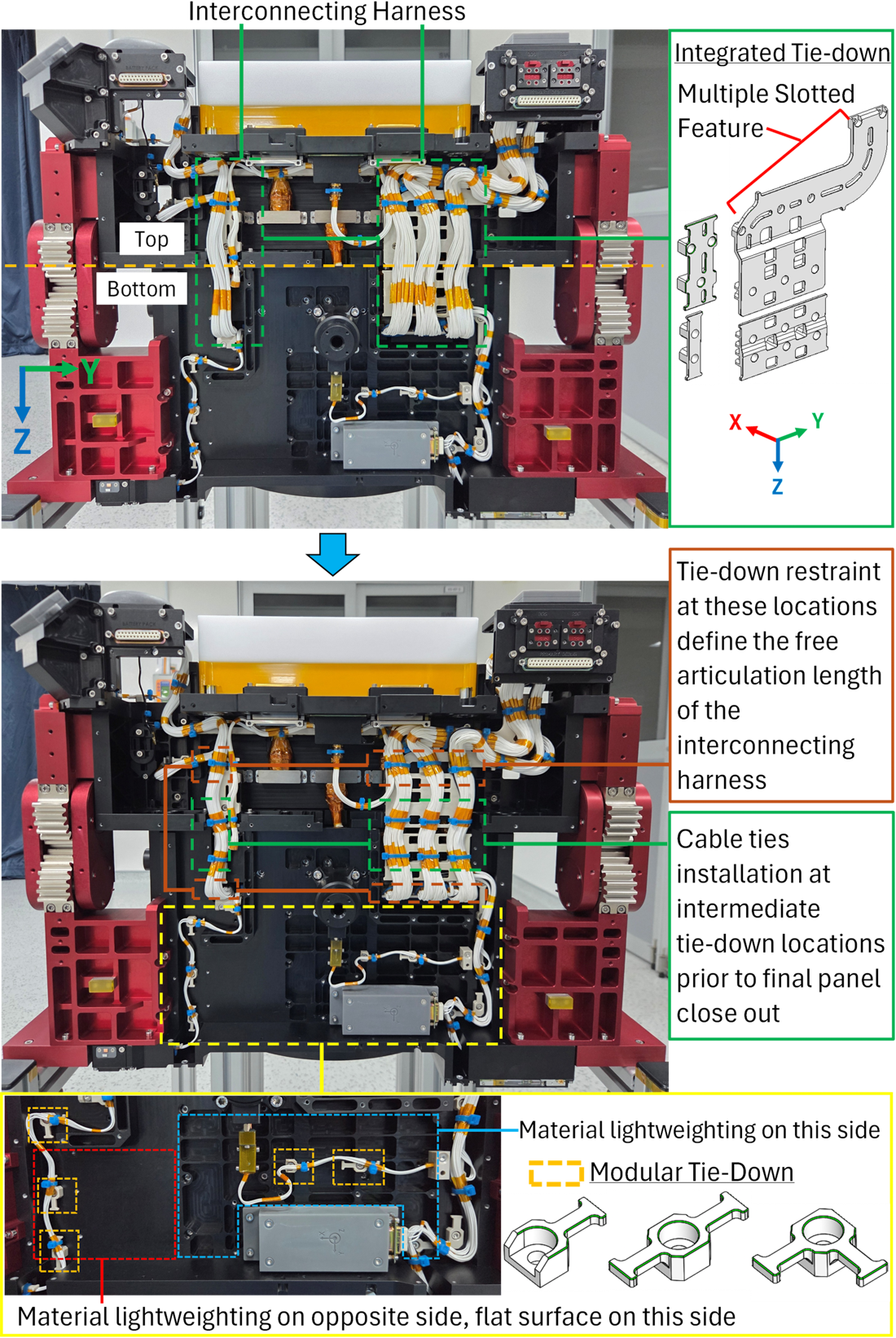}\\
\caption{Integrated and modular PEEK harness tie-down brackets implemented on the Lumelite-5 $-X$ side.}
\label{fig:fig11}
\end{figure}
\begin{figure}
\centering
\includegraphics[width=3.25in]{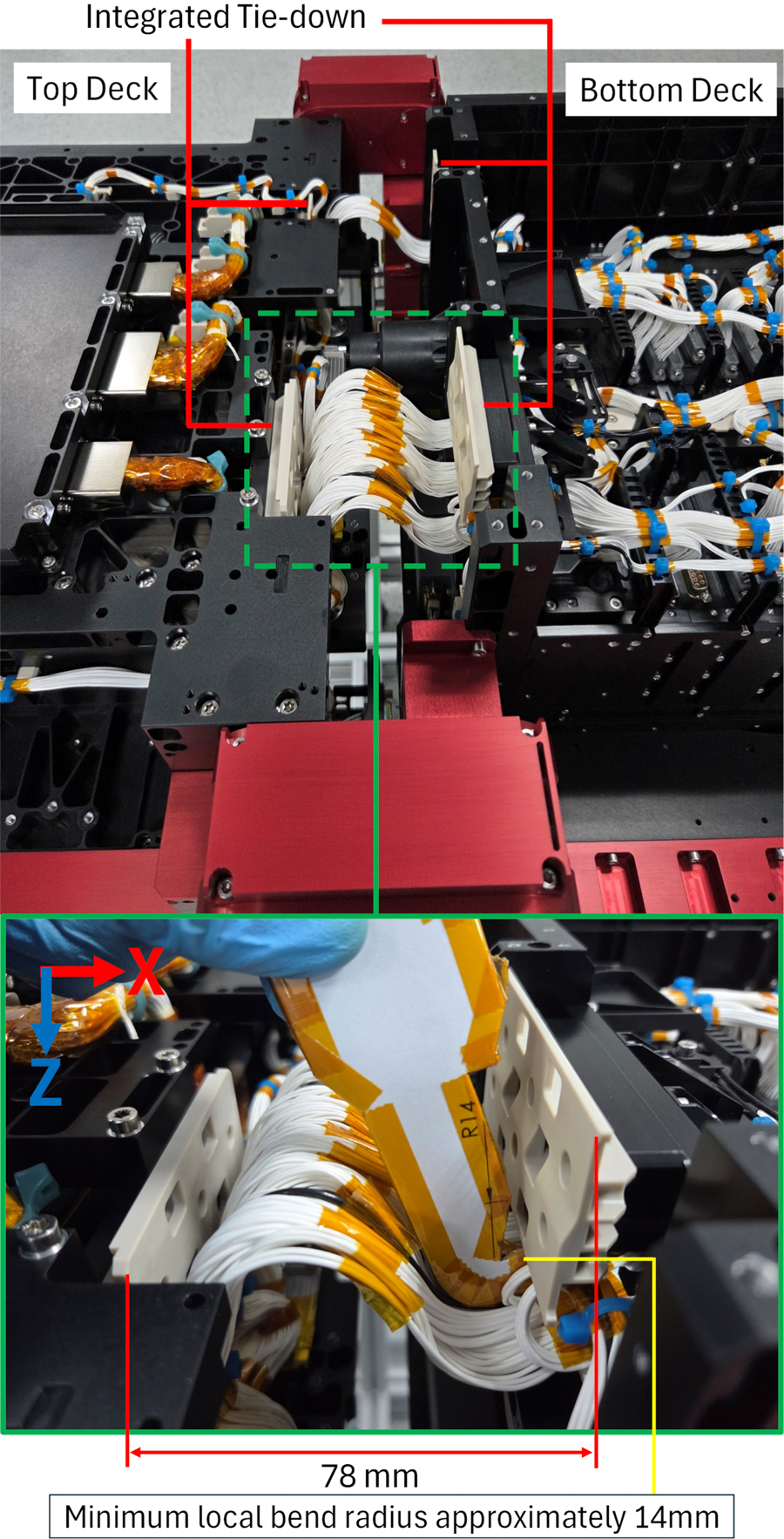}\\
\caption{Interconnecting-harness configuration with the top deck opened to 180$^\circ$ and radius-template assessment of the minimum local bend radius.}
\label{fig:fig12}
\end{figure}
Harness routing and tie-down locations were developed in parallel with the satellite architecture, subsystem and component placement, and MGSE design. Harness tie-down retainers are mechanically fastened to the satellite structure or subsystem mounting surfaces rather than adhesively bonded. Figure~\ref{fig:fig10} shows one of the examples of aluminum-alloy tie-down retainers integrated on bus electronics that are used where concentrated routing, limited packaging space, and thin-wall or geometrically complex features are required. In other regions, integrated and modular PEEK tie-down retainers are used, as shown in Figure~\ref{fig:fig11}. PEEK was selected primarily for lightweighting, while the modular tie-down retainers share a common mounting interface that permit different retainer configuration to be installed at the same structural interface. Multiple slot features are incorporated into the retainers to allow local adjustment of the harness-restraint position. This accommodates differences between idealized harness geometry represented in CAD and the as-built configuration resulting from assembly and manufacturing tolerances.

The tie-down arrangement is also coordinated with the double-hinge MGSE. While top-deck rotation is required, intermediate tie-down locations along the articulating harness section are intentionally left without cable-tie restraints. These intermediate restraints are installed only prior to final satellite closeout, after double-hinge operation is no longer required. The remaining tie-down restraint locations on the top and bottom decks define the free articulation length of the interconnecting harness, allowing the harness section between these points to articulate during top-deck rotation, as illustrated in Figure~\ref{fig:fig11}.

Figure~\ref{fig:fig12} shows the interconnecting harness configuration with the top deck opened to 180$^\circ$. Using verified printed radius templates, the smallest local bend radius of the articulated harness was estimated to be approximately 14~mm. The interconnecting harness includes Times Microwave Systems 52091 RF cables connecting the bus electronics on the bottom deck to the GPS receiver and S-band transceiver on the top deck towers. These cables specify a nominal dynamic minimum bend radius of 0.437~in., corresponding to approximately 11.1~mm \cite{rfcable}. This bend-radius requirement was therefore considered when defining the restraint locations and free articulation length of the interconnecting harness.

The AIT benefits and implementation value of the co-design framework, incorporating the aforementioned Techniques 1–4, are discussed in detail next.

\section{AIT Benefits and Implementation Value}
\subsection{Technique 1: Centralized Service Interfaces}
Champlin described the Iridium design-for-testability approach, in which a dedicated test port provided direct access to the spacecraft digital circuitry and onboard computers during production testing, while built-in test capability supported testing at multiple levels of integration \cite{champlin_iridium}. Baghal later summarized several Iridium lessons applicable to rapid spacecraft production, including designing for repair and upgrades, test automation, built-in and self-test capability, and appropriately located test interfaces; the same work reported a production rate of approximately 22 days per spacecraft \cite{baghal_pnpsat-1}.

For Lumelite-5, accessibility during thermal-vacuum testing and launch-vehicle upper-stage integration was the primary driver for centralizing the bus and payload service interfaces together with the battery charging interface. During thermal-vacuum testing, access is typically greatest on the satellite side facing the chamber door, while access to the remaining sides may be constrained by the chamber size and test setup. Similarly, during launch-vehicle integration, neighboring satellites or surrounding integration hardware can restrict access to portions of the satellite. Centralizing the interfaces within a common accessible region therefore facilitates EGSE setup and checkout during thermal-vacuum testing and reduces the risk of service and battery charging interfaces becoming obstructed during launch-site integration.

Centralization also provides secondary AIT benefits by reducing the number of satellite sides that operators need to access during test preparation, which may reduce the likelihood of missed or incorrectly connected interfaces. This approach is consistent with Baghal's recommendation that rapid-AIT facilities be designed to reduce hazardous and time-consuming spacecraft movement, minimize GSE movement, and reduce the time between operations \cite{baghal_pnpsat-1}. The common-access arrangement is also beneficial when testing is performed at external facilities with constrained working space, as most service and charging connections can be established from one region without requiring repeated movement around the satellite and surrounding test equipment.
\subsection {Technique 2: Skeletal Support Interfaces}
Previous small-satellite integration experience has shown that internal access can become increasingly restricted as panel integration progresses, making assembly sequence and harness management important AIT considerations \cite{kanji_norsat-1,gwozdecky_defiant}. The Aspera DEFIANT-XL integration further highlighted harness-handling concerns, where overhanging harnesses were at risk of snagging underneath a tray edge during spacecraft assembly \cite{leaf_mgse}.

\begin{figure}
\centering
\includegraphics[width=3.25in]{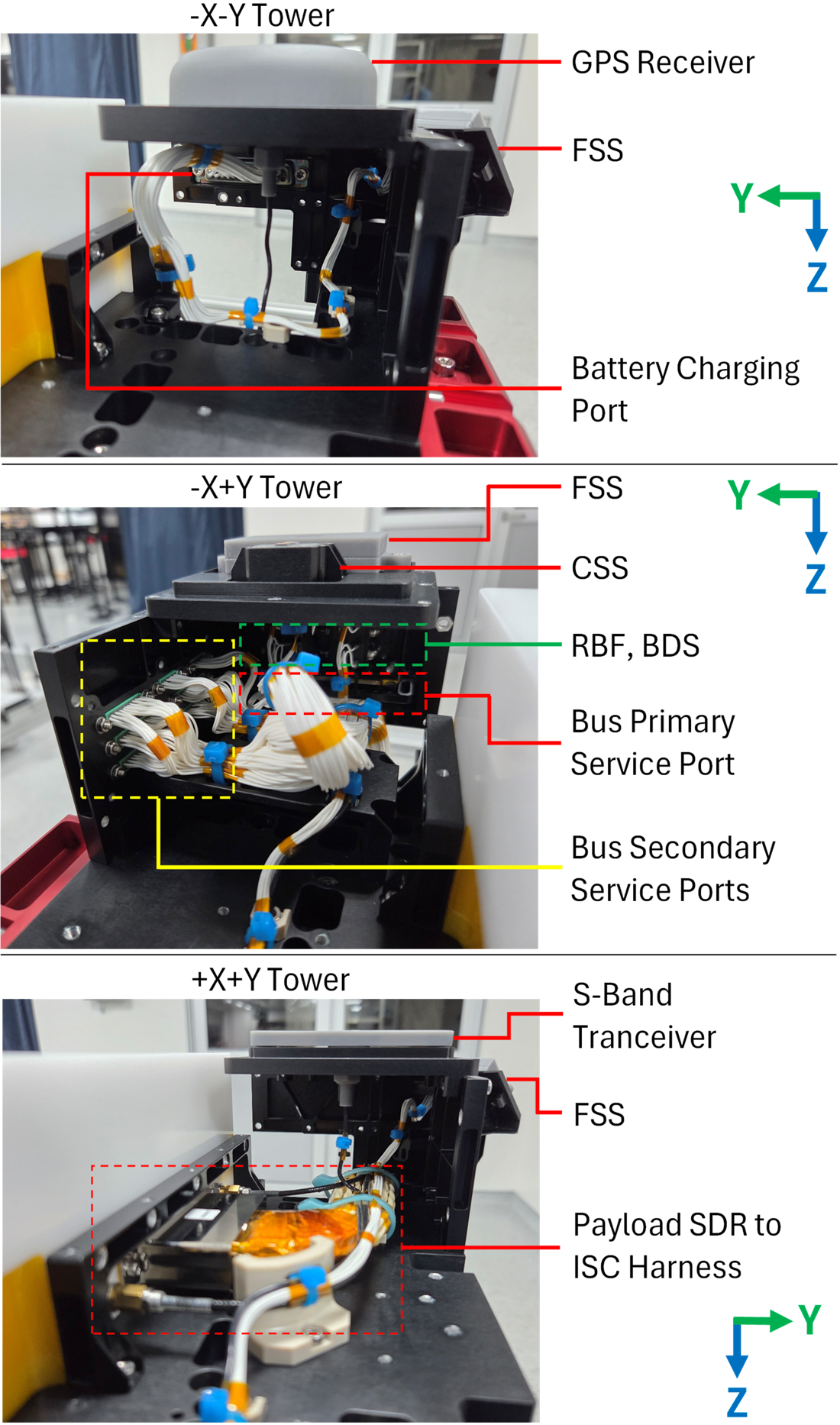}\\
\caption{$-X-Y$, $-X+Y$, and $+X+Y$ skeletal support towers with the enclosures removed, exposing internal harness routing and tie-down points.}
\label{fig:fig13}
\end{figure}

In Lumelite-5, removal of the tower enclosures exposes the skeletal support structures and provides direct access for component installation, connector mating, harness routing, and tie-down activities, as shown in Figure~\ref{fig:fig13}. Components and service interfaces remain mounted on the skeletal structures and can therefore be integrated before enclosure closeout without the access restrictions that would otherwise be introduced by the surrounding enclosures.

After integration, the removable enclosures can be removed and reinstalled without disconnecting the components and service interfaces or disturbing the established harness configuration. This reduces unnecessary component handling, additional connector mating cycles, loose-harness handling, and the assembly and disassembly steps required to gain internal access for inspection, troubleshooting, or rework. If removal of a sensor or component bracket is required, the defined mechanical references described in Section~2 provide controlled positioning for subsequent reinstallation, supporting repeatable bracket placement and reducing reliance on manual realignment.
\subsection {Technique 3: Double-Hinge MGSE}
\begin{figure}
\centering
\includegraphics[width=3.25in]{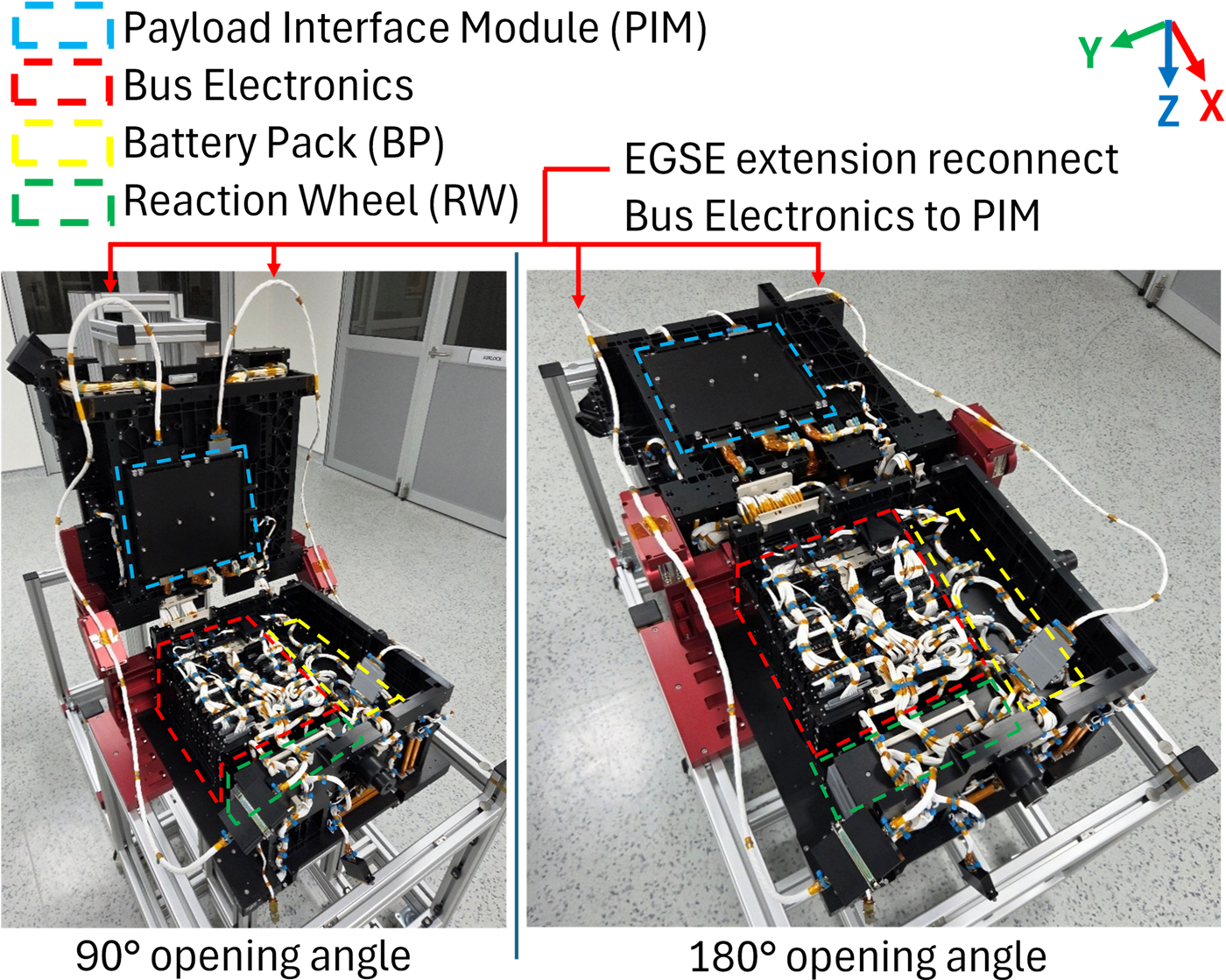}\\
\caption{Top deck opened to 90$^\circ$ (left) and 180$^\circ$ (right), providing access to bottom-deck bus subsystems and components.}
\label{fig:fig14}
\end{figure}
The double-hinge MGSE provides direct access to the bottom-deck subsystems and components during both initial assembly and subsequent integration activities. During initial assembly, opening the top deck to 90$^\circ$ or 180$^\circ$ provides a large working envelope for component installation, connector mating, harness routing, inspection, and other integration activities on the bottom deck. This reduces the access restrictions that would otherwise arise if these activities were performed with the top deck in its closed configuration.

The same accessibility is retained as satellite integration progresses. Figure~\ref{fig:fig14} shows the top deck opened to 90$^\circ$ and 180$^\circ$, providing access to bottom-deck bus subsystems and components even after the satellite has been substantially integrated. This allows inspection, troubleshooting, component replacement, and rework without extensive disassembly of the satellite structure. The 180$^\circ$ configuration further provides a larger working envelope around the bottom deck when greater access is required. Such flexibility is particularly useful when the actual integration sequence deviates from the planned procedure. Gwozdecky reported that the first flight build rarely proceeded in the intended integration order, with Mission 2 spacecraft frequently built out of order to accommodate missing parts or schedule drivers \cite{gwozdecky_defiant}. 

In addition to providing physical access, the double-hinge MGSE allows most bus-level functionality to be retained while the top deck remains open. With the EGSE extension described in Section~2, the majority of the satellite subsystems remain electrically connected in the open configuration. Functional checks, troubleshooting, burn-in activities, and staged build-and-test activities can therefore be performed while the relevant hardware remains physically accessible. This is particularly useful when a fault is identified because the affected hardware can be operated and diagnosed without first returning the satellite to its closed configuration. The MGSE can subsequently be removed and the solar panels reinstalled when required for testing.

Implementing the mechanism as MGSE rather than incorporating permanent hinges into the flight structure also transfers the associated hinge mechanism, support hardware, mass, and packaging volume to the ground-support equipment. The satellite therefore gains the integration-accessibility function without carrying the corresponding mechanism as flight hardware.
\subsection {Technique 4: Integrated and Modular Tie-Down Retainers}
Conventional adhesively bonded harness tie-down retainers require surface preparation, adhesive handling, curing time, and a sufficiently large flat bonding surface on the supporting structure. These requirements can become restrictive on lightweighted structural panels, where pockets and ribs reduce the available bonding area and may also constrain the harness-routing path. Figure~\ref{fig:fig11} illustrates this constraint on the Lumelite-5 $-X$ structural panel. In the red-dotted region, bonded tie-down retainers remain feasible because a sufficiently large flat surface is available, with the material-lightweighting features located on the opposite face of the panel. In contrast, the blue-dotted region contains lightweighting features on the exposed panel surface, where the resulting pockets and ribs restrict the available flat bonding area and make installation of bonded tie-down retainers impractical. The lightweighting arrangement is influenced by the mounting of the bus electronics on the opposite side of the panel, where sufficient surface contact must be retained to provide a conductive thermal path to the structural panel. To overcome the limitation of bonded retainers, mechanically fastened tie-down retainers were implemented on Lumelite-5. Unlike bonded retainers, these mechanically fastened retainers can be integrated at predefined mounting interfaces on both regular flat surface and lightweighted regions.

Gwozdecky observed that flight harness routing can differ from that developed using earlier jigs or representative configurations, and that tie-down locations may consequently need to change or be introduced at a later stage \cite{gwozdecky_defiant}. To accommodate differences between the idealized harness geometry represented in CAD and the as-built configuration, the integrated tie-down retainers implemented on Lumelite-5 incorporate multiple slots that allow local adjustment of the harness-restraint position. Where additional local routing adjustment is required, modular PEEK tie-down retainers with common mounting interfaces can be interchanged without modifying the underlying structure.

\begin{table*}[!t]
\centering
\caption{Summary of Lumelite-5 Co-Design Implementation and AIT Value}
\label{tab:codesign_summary}
\renewcommand{\arraystretch}{1.25}
\footnotesize

\begin{tabularx}{\textwidth}{
    >{\raggedright\arraybackslash}p{0.18\textwidth}
    >{\raggedright\arraybackslash}p{0.30\textwidth}
    >{\raggedright\arraybackslash}X}
\hline
\textbf{Technique / Framework Level} &
\textbf{Implemented Feature} &
\textbf{Implementation Evidence / AIT Value} \\
\hline

Centralized service interfaces &
Primary, secondary, and payload service ports, RBF, BDS, and the separate
battery charging interface centralized on the $-X$ side &
Maintains a common accessible region for EGSE checkout and battery charging
during thermal-vacuum testing and launch-vehicle integration, where access to
other satellite sides may be constrained. \\
\hline

Skeletal support interfaces &
Removable enclosures retain components and service interfaces on the
skeletal structures; defined mechanical references are incorporated for
sensor and component positioning &
Preserves access to components, connectors, and harness routing during and
after integration while supporting controlled removal and reinstallation of
sensor and component brackets during inspection, troubleshooting, or rework. \\
\hline

Double-hinge MGSE &
Top deck can be opened and supported at 90$^\circ$ and 180$^\circ$; an EGSE
extension reconnects the bus electronics to the PIM in the open configuration &
Provides access to bottom-deck subsystems and components after substantial
satellite integration while allowing the majority of satellite subsystems to
remain electrically connected for functional activities. \\
\hline

Integrated and modular tie-down retainers &
Mechanically fastened retainers with adjustable slots, interchangeable modular
configurations, and predefined top- and bottom-deck restraint locations &
Avoids bonding-surface and adhesive-process constraints, accommodates
as-built routing variation, and provides controlled harness articulation;
the approximately 14~mm minimum local bend radius at 180$^\circ$ remains
above the 11.1~mm nominal dynamic minimum specified for the applicable
RF cable. \\
\hline

Combined framework &
Four techniques implemented as coordinated satellite, harness, MGSE, and
AIT provisions &
Preserves physical accessibility, electrical connectivity, and harness control
through progressive integration, supporting staged functional verification,
troubleshooting, rework, and final inspection while minimizing unnecessary
disturbance to integrated hardware. \\
\hline

\end{tabularx}
\end{table*}

For the interconnecting harnesses involved in double-hinge operation, the predefined restraint locations on the top and bottom decks provide a controlled articulation region during top-deck rotation. By establishing the free harness length between these restraint points, the arrangement maintains sufficient slack while limiting local bending as the top deck is opened. The approximately 14~mm minimum local bend radius estimated at the 180$^\circ$ configuration remains above the 11.1~mm nominal dynamic minimum specified for the applicable RF cables \cite{rfcable}. The tie-down arrangement therefore allows the interconnecting harnesses to accommodate changes in satellite top deck opening angle while preserving the established routing configuration.

More generally, the predefined tie-down points provide a structured routing framework throughout assembly and integration. They provide physical references for establishing and restoring harness routing during installation, removal, or rework while allowing local adjustment where required by the as-built configuration.
\subsection {Integrated Co-Design Framework for AIT}
The value of the proposed approach arises from the interaction of the four co-designed elements rather than from any single feature in isolation. The centralized service interfaces maintain accessible electrical connections for checkout during AIT and launch vehicle integration; the skeletal support interfaces provide access for component installation and harness routing on the top deck during initial assembly while preserving access after integration; the double-hinge MGSE provides similar access to bottom-deck subsystems and components during both initial assembly and subsequent integration activities; and the dedicated harness-routing and tie-down provisions provide a structured and adjustable routing framework as the satellite configuration evolves during AIT. Together, these features preserve physical accessibility, electrical connectivity, and harness control throughout progressive satellite integration.

The skeletal support interfaces and double-hinge MGSE complement one another by maintaining access to different satellite regions. With the tower enclosures removed, components, connectors, harness-routing paths, and tie-down locations on the top deck remain accessible for initial installation, inspection, and rework. Similarly, opening the top deck to 90$^\circ$ or 180$^\circ$ provides direct access to the bottom-deck integration region. Access is therefore retained not only during initial assembly but also after the satellite has become substantially integrated, reducing the need to undo completed integration work when troubleshooting, component replacement, or corrective action is required.

Physical accessibility is complemented by retained electrical connectivity. While the skeletal support interfaces and double-hinge MGSE expose the relevant integration regions, the service interfaces and EGSE connections allow most bus-level functionality to remain available for functional activities. The satellite can therefore be operated, inspected, and diagnosed while the relevant hardware remains physically accessible, linking mechanical access and electrical testability within the same integration configuration.

The harness-routing and tie-down provisions support this accessibility by maintaining controlled routing as the satellite configuration changes. In particular, the defined restraint locations allow the interconnecting harnesses to articulate during double-hinge operation without requiring the established routing configuration to be reconstructed after each opening cycle. The harness provisions therefore work together with the skeletal support interfaces and double-hinge MGSE to preserve both access and harness control as integration progresses.

The combined framework also facilitates burn-in and staged build-and-test activities. Burn-in is intended to detect infant or early-life failures \cite{teo}, while intermediate bus functional tests performed during assembly can identify faulty components early in the assembly process before further integration activities are completed \cite{baghal_pnpsat-1}. In Lumelite-5, accessible service interfaces, removable enclosures, an openable top deck, and controlled harness routing allow subsystems and components to be progressively integrated and functionally verified while access is retained for troubleshooting, replacement, or rework.

Early fault detection is particularly important because faults discovered only after environmental testing may require additional disassembly, corrective work, re-integration, and potentially repeated environmental testing, depending on the extent of the rework. The resulting schedule impact can be greater when testing is performed at external facilities, where access to shaker tables, thermal-vacuum chambers, and other specialized equipment depends on facility availability and scheduling. Repeated environmental testing may therefore introduce additional test duration as well as facility rebooking, transportation, setup, and post-test activities.

Finally, the integrated framework preserves access before final satellite closeout. With the tower enclosures removed and the top deck opened using the double-hinge MGSE, the principal integration regions, connectors, harness-routing paths, tie-down points, and mechanical interfaces remain accessible for final inspection and corrective work. The combined framework therefore supports progressive assembly, intermediate functional verification, troubleshooting and rework, and final inspection while minimizing unnecessary disturbance to hardware that has already been integrated. Table~\ref{tab:codesign_summary} summarizes the implemented co-design elements, their principal features, and the corresponding implementation evidence and AIT value discussed for the Lumelite-5 SM.

\begin{figure}[!t]
\centering
\includegraphics[width=3.25in]{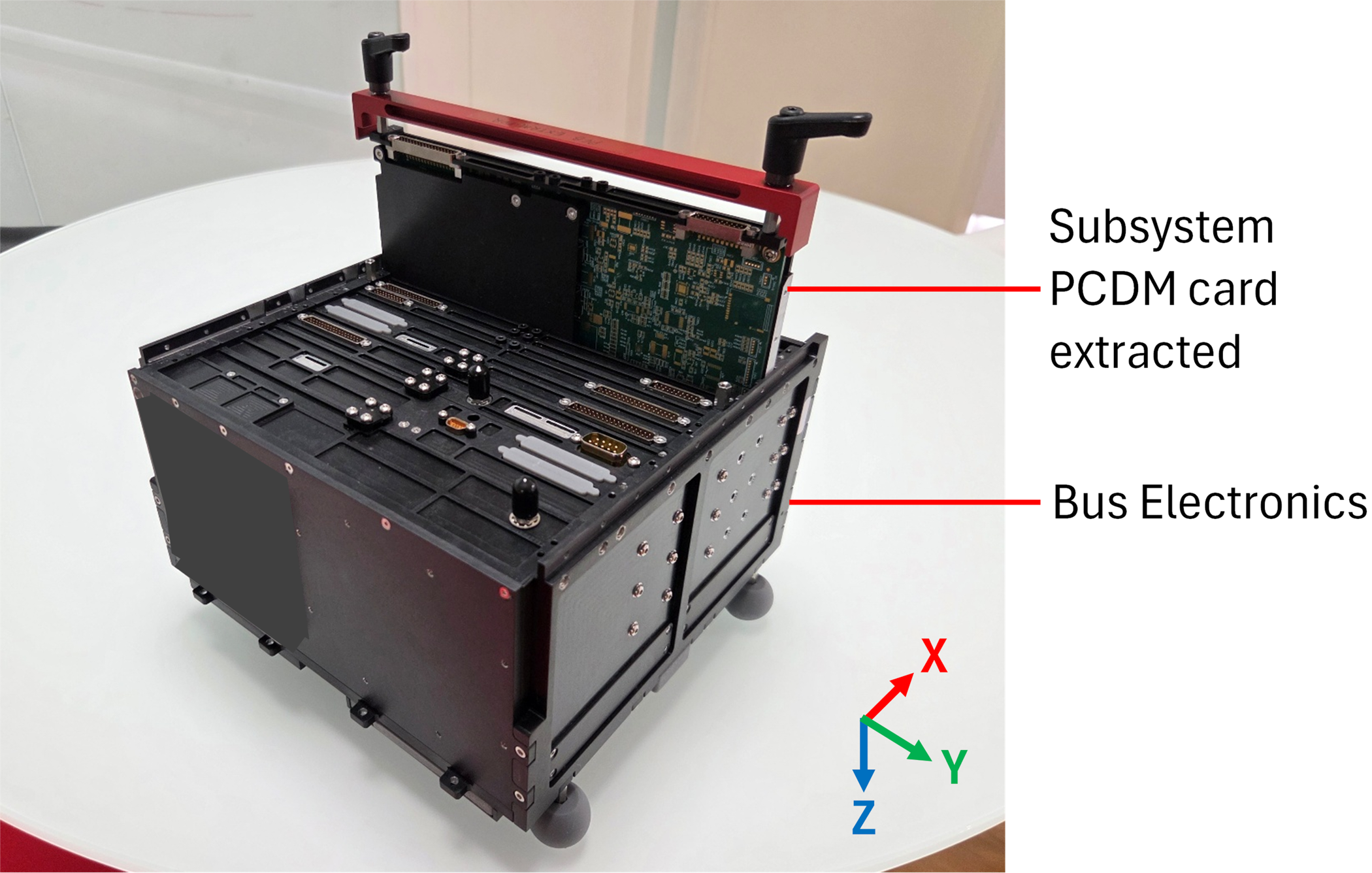}\\
\caption{Bus electronics assembly with one subsystem card extracted, illustrating the card-level removability of the modular bus electronics design.}
\label{fig:fig15}
\end{figure}
\section{Limitations and Future Improvements}
\subsection{Limitations of the Present Lumelite-5 Implementation}
Although the integrated co-design framework improves physical accessibility and supports AIT activities throughout multiple stages of integration, several design and evaluation limitations remain in the present Lumelite-5 implementation. These limitations also provide useful design feedback for subsequent satellite-bus developments.

The present evaluation is based primarily on SM integration activities, and the complete Lumelite-5 test campaign has not yet been completed. The implementation value is therefore assessed mainly through physical accessibility, interface behavior, harness routing, and integration workflow rather than through a controlled before-and-after comparison of AIT cycle time or manpower. Accordingly, this paper does not claim a quantified reduction in total AIT duration or labor.

Besides, the first design limitation concerns the SDR payload located on the top deck. Once the top deck is integrated, the SDR payload is no longer directly accessible without substantial disassembly. Electrical access remains available through the payload service interfaces, enabling functional checkout and troubleshooting via EGSE, but direct access to the payload hardware is restricted. No redesign is pursued within the present Lumelite-5 program to address this limitation, and it therefore remains a trade-off of the current satellite configuration.

The second design limitation concerns harness routing above the bus electronics on the bottom deck. The bus electronics employ a modular architecture in which individual subsystem cards are connected to a backplane PCB and can, in principle, be extracted independently from the enclosure without dismantling the complete bus-electronics assembly. Figure~\ref{fig:fig15} shows one subsystem card extracted from the bus electronics. However, in the current configuration, 24 harness assemblies originate from the bus electronics, of which 22 are routed above the bus-electronics enclosure. Consequently, extraction of an individual subsystem card may require substantial disassembly and handling of the overlying harnesses before sufficient access is obtained. This reduces the practical benefit of the modular card architecture during troubleshooting, replacement, and rework.

This routing limitation is partly driven by the satellite-level interface architecture. Accessibility during thermal-vacuum testing and launch-vehicle upper-stage integration drove the requirement to centralize the primary, secondary, and payload service interfaces together with the battery charging interface within a common accessible region. In the present Lumelite-5 configuration, four independently deployable solar panels occupy the satellite sides, and the centralized interfaces are therefore located on the top deck to maintain accessibility when the panels are stowed. The corresponding harnesses must consequently be routed from the bottom deck to the top deck, contributing both to the harness concentration above the bus electronics and to the number of interconnecting harnesses between the two decks.

\begin{figure}[!t]
\centering
\includegraphics[width=3.25in]{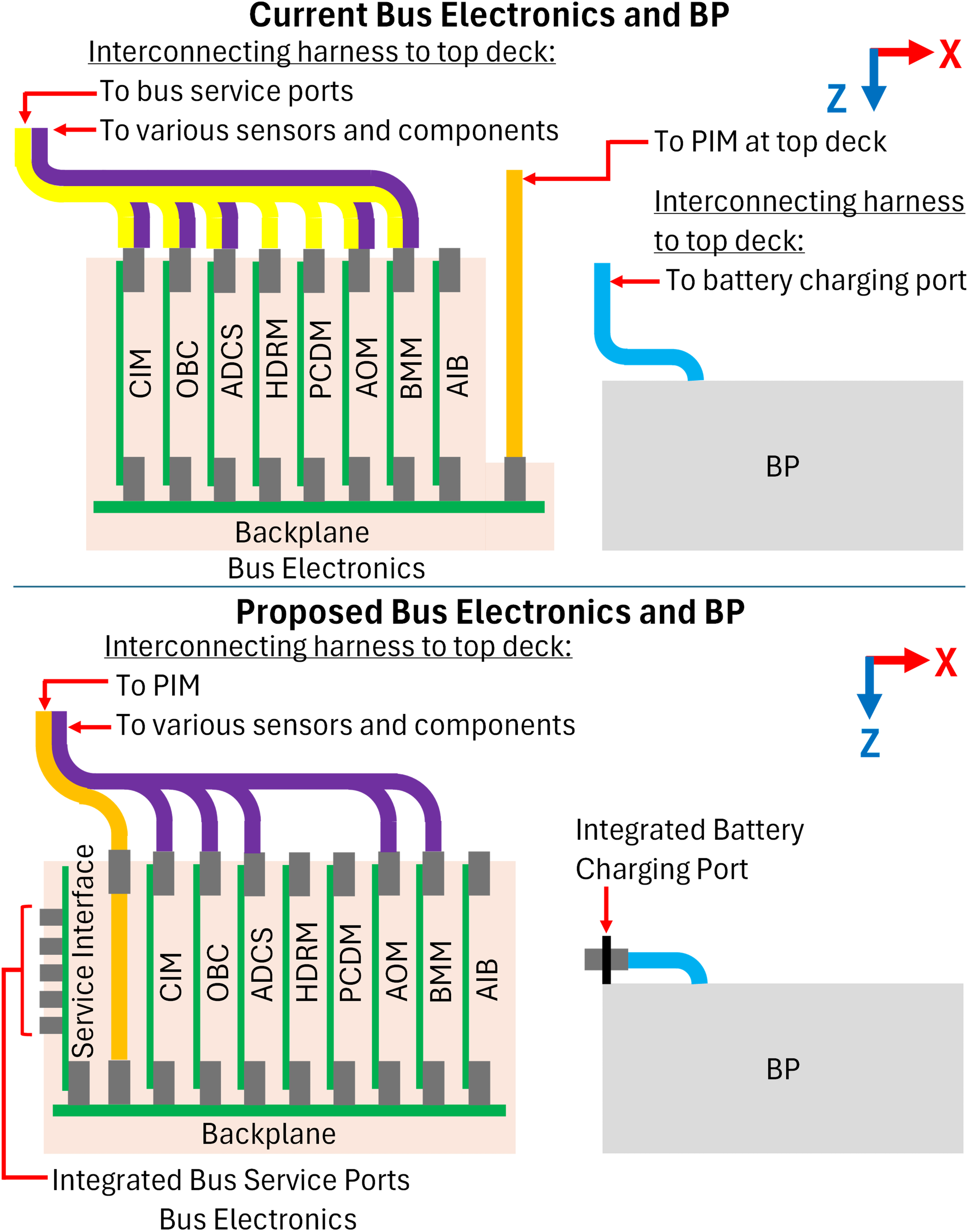}\\
\caption{Conceptual reconfiguration of the bus electronics and battery pack, showing the proposed integration of the service interfaces into the bus electronics and the separate battery charging port into the battery pack.}
\label{fig:fig16}
\end{figure}

\subsection{Conceptual Satellite Reconfiguration}
\begin{figure}
\centering
\includegraphics[width=3.25in]{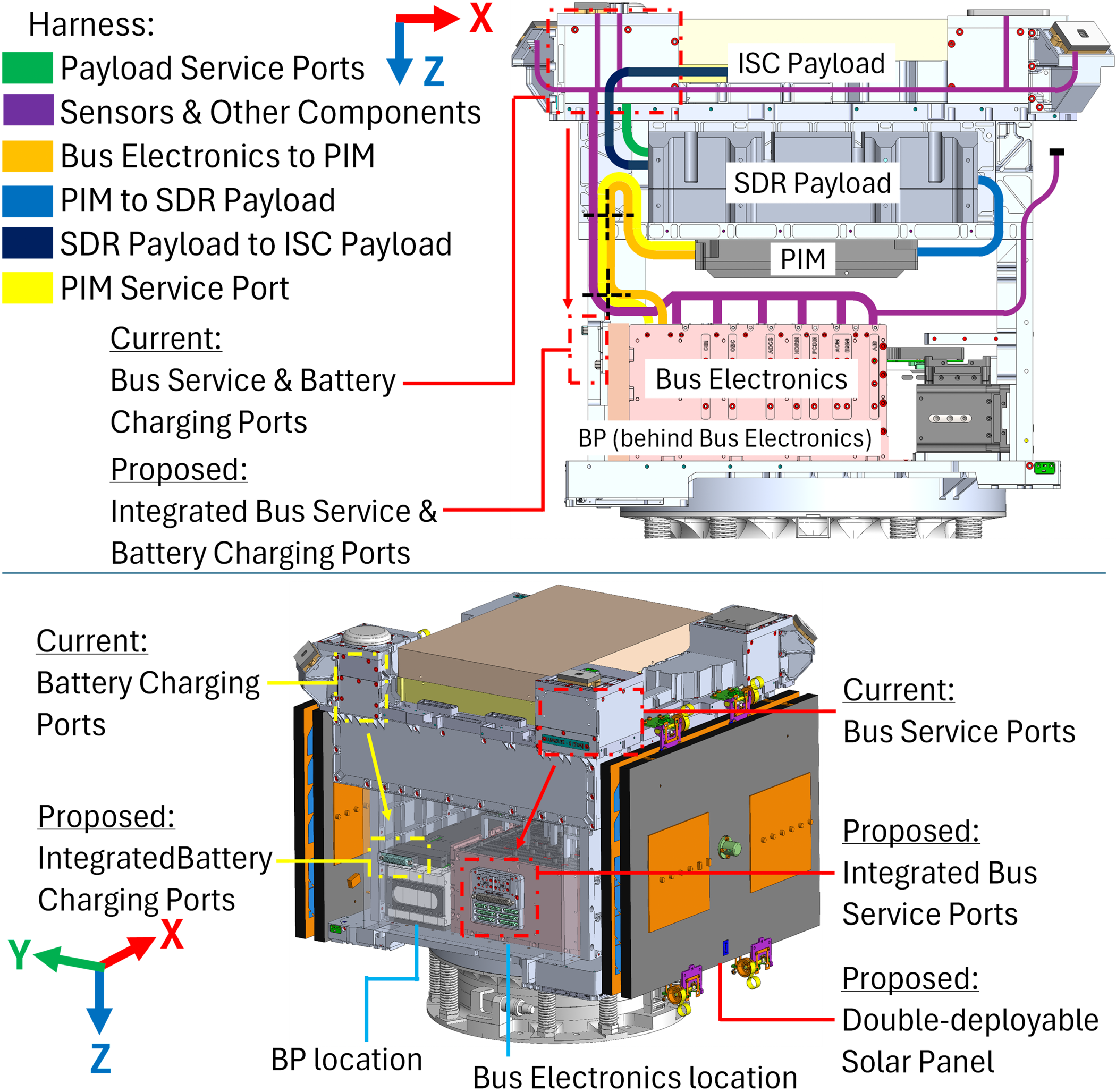}\\
\caption{Conceptual satellite-level reconfiguration with two double-deployable solar panels, showing reduced harness routing and continued accessibility of the service interfaces on the satellite $-X$ side in the stowed configuration.}
\label{fig:fig17}
\end{figure}

A conceptual study was therefore performed to investigate an alternative satellite configuration that could reduce the harness-routing burden above the bus electronics while preserving centralized interface accessibility during thermal-vacuum testing and launch-vehicle integration. The proposed configuration replaces the four independently deployable solar panels with two double-deployable solar panels, providing sufficient access on the satellite $-X$ side for the service interfaces to be integrated directly into the bus electronics while remaining accessible with the solar panels stowed. As shown in Figure~\ref{fig:fig16}, the proposed bus-electronics architecture incorporates a dedicated service-interface PCB card that interfaces with the relevant subsystem cards through the backplane PCB. The battery charging port is separately integrated directly into the battery pack (BP). Figure~\ref{fig:fig17} shows the resulting satellite-level configuration, in which the centralized interfaces remain accessible from the $-X$ side with the solar panels stowed.

The proposed configuration would reduce the number of harness assemblies routed above the bus electronics from 22 to 12. The number of interconnecting harness assemblies routed between the top and bottom decks through the double-hinge MGSE articulation region would also decrease from 16 to 8. A total of 10 bus-electronics-to-service-port harnesses would be eliminated, while the PIM service harness would become the only remaining service harness crossing between the two decks. The battery-charging harness between the bottom and top decks would also be eliminated by integrating the charging interface directly into the BP.

\subsection{Expected AIT Improvements and Design Implications}
The proposed routing would also simplify operation of the double-hinge MGSE. In the current configuration, the two bus-electronics-to-PIM connectors must be disconnected prior to top-deck rotation. At the 90$^\circ$ and 180$^\circ$ configurations, an EGSE extension harness is then required to reconnect the bus electronics to the PIM so that the majority of the satellite subsystems remain electrically connected while the top deck is open. In the proposed configuration, the two bus-electronics-to-PIM harnesses would instead be routed through the double-hinge MGSE articulation region and would remain connected throughout top-deck rotation. The majority of the satellite subsystems could therefore remain electrically connected in the open configuration without requiring the EGSE extension harness. The $+X$ solar-panel bulkhead connector would become the only connector requiring disconnection prior to top-deck rotation, reducing the required connector disconnections from 3 to 1. Table~\ref{tab:table3} summarizes the corresponding harness-assembly counts and connector-disconnection requirements.

\begin{table}[!t]
\centering
\caption{Summary of the conceptual reconfiguration comparing harness-assembly counts and connector disconnections.}
\label{tab:table3}
\renewcommand{\arraystretch}{1.15}
\footnotesize

\begin{tabularx}{\columnwidth}{
    >{\raggedright\arraybackslash}X
    >{\centering\arraybackslash}p{0.16\columnwidth}
    >{\centering\arraybackslash}p{0.16\columnwidth}}
\hline
\textbf{Item} &
\textbf{Current} &
\textbf{Proposed} \\
\hline

\multicolumn{3}{l}{\textit{Harness assemblies routed above bus electronics}} \\
Harness assemblies &
22 &
12 \\
\hline

\multicolumn{3}{l}{\textit{Interconnecting harnesses through double-hinge MGSE}} \\
Bus electronics to various sensors and components &
5 &
5 \\

Bus electronics to service ports &
10 &
0 \\

PIM to service port &
0 &
1 \\

BP to battery charging port &
1 &
0 \\

Bus electronics to PIM &
0 &
2 \\

\textbf{Total} &
\textbf{16} &
\textbf{8} \\
\hline

\multicolumn{3}{l}{\textit{Connectors requiring disconnection prior to top-deck rotation}} \\
Bus electronics to PIM &
2 &
0 \\

Bus electronics to $+X$ solar panel &
1 &
1 \\

\textbf{Total} &
\textbf{3} &
\textbf{1} \\
\hline

\end{tabularx}
\end{table}
The proposed changes would eliminate 10 service harnesses and 1 battery-charging harness presently routed between the bottom and top decks. These 11 harnesses range from approximately 342~mm to 413~mm in length. Their elimination would also allow the associated tie-down provisions to be reduced, resulting in a less congested routing arrangement above the bus electronics and between the two decks, as illustrated conceptually in Figure~\ref{fig:fig17}. In addition to reducing harness-handling requirements during subsystem-card removal, maintaining the bus-electronics-to-PIM connections throughout top-deck rotation would remove the additional disconnection and reconnection steps presently required to restore most bus-level functionality.

The proposed reconfiguration would require broader changes to the satellite architecture, including the solar-panel configuration, connector locations, PCB layout, structural packaging, and harness-routing scheme. These impacts have not been assessed in detail; the concept should therefore be regarded as a preliminary improvement study and will not be implemented within the current program. Nevertheless, it provides a basis for future satellite-bus development by illustrating how early decisions concerning solar-panel architecture, interface placement, subsystem modularity, and harness routing can be co-developed to improve physical accessibility and reduce AIT handling requirements.

\section{Extensions Towards Autonomous AIT}

Recent work has demonstrated the potential of cyber-physical systems and extended-reality-assisted workflows to support aerospace AIT through structured assembly instructions, object recognition and tracking, and real-time operator guidance \cite{marinacci}. These approaches complement the physical design-for-AIT provisions considered in the present work and suggest a broader pathway toward increasingly integrated human-assisted and automated AIT workflows. In this context, the implemented Lumelite-5 framework and the conceptual improvements identified in this study provide a basis for progressively increasing AIT automation in future satellite programs. Building on STAR's prior implementation of automation-assisted test and checkout activities on Lumelite-4, future developments could combine automated electrical verification with a satellite architecture designed to provide controlled physical access, defined mechanical interfaces, structured harness routing, and accessible service connections. Together, these features could provide more clearly defined and repeatable interfaces for both human operators and automated equipment.

\begin{figure}[!t]
\centering
\includegraphics[width=3.25in]{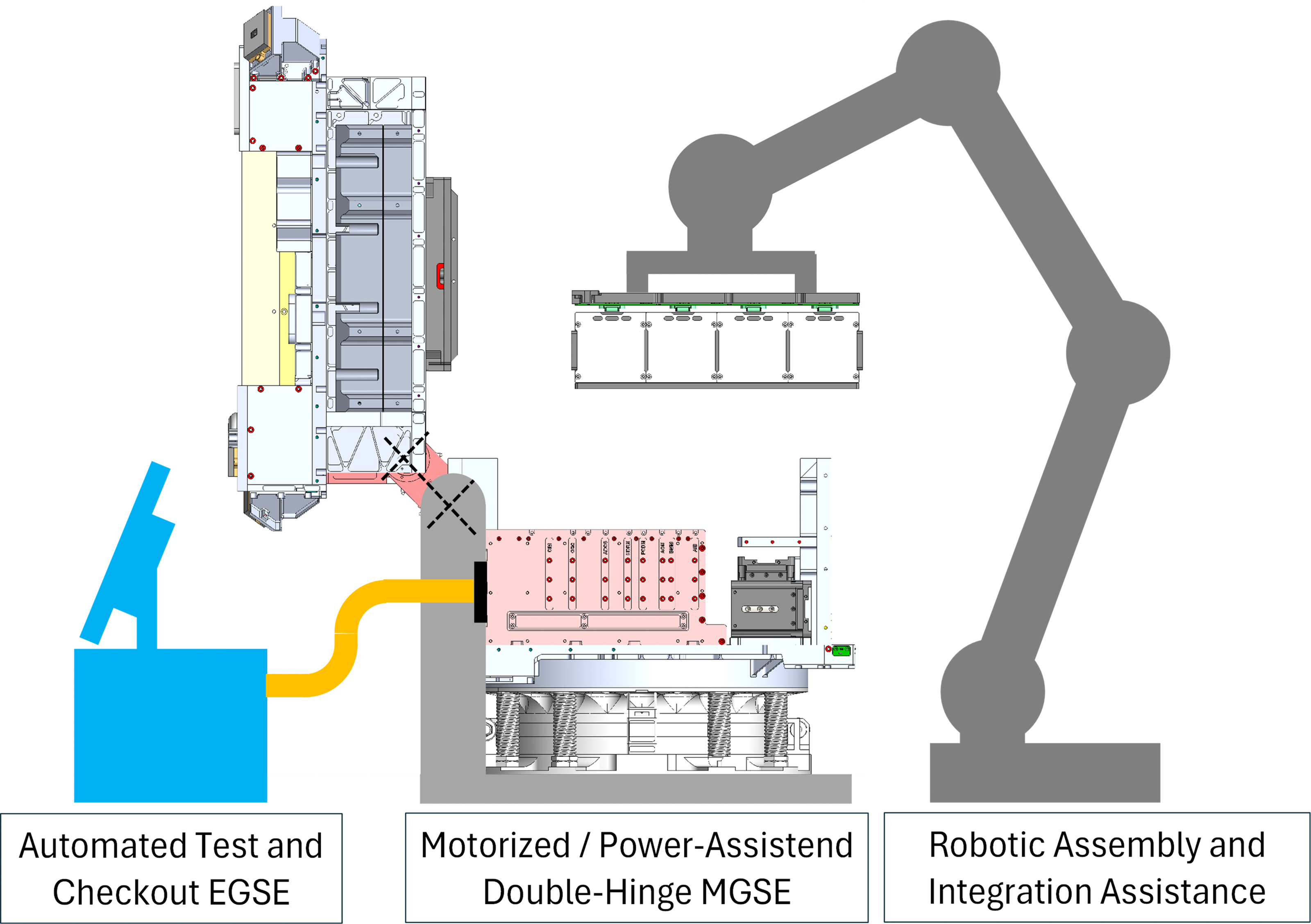}\\
\caption{Conceptual future AIT architecture combining automated test and checkout EGSE, motorized or power-assisted double-hinge MGSE, and robotic assistance for human-in-the-loop satellite assembly and integration.}
\label{fig:fig18}
\end{figure}

The double-hinge MGSE provides one example of how the present framework could be extended toward assisted or automated integration. A future implementation could motorize or power-assist the hinge mechanism to control top-deck opening, closing, and positioning. Combined with robotic or other automated handling equipment, such an approach could assist repetitive and physically demanding activities including lifting, positioning and placing subsystem assemblies. Recent work on automated satellite assembly has demonstrated the use of feature holes on satellite mating interfaces as assembly and measurement reference points for high-precision component pose measurement \cite{zhang}, providing a relevant example of how defined interface features can support automated positioning and assembly.

Within a future Lumelite-5-derived AIT architecture, automated equipment could undertake or assist these handling-intensive activities, while human operators focus on tasks requiring greater dexterity and engineering judgement, including harness integration, inspection, troubleshooting, and corrective work. Automated test and checkout EGSE could also be integrated into this workflow to support progressive functional verification as satellite assembly proceeds. Figure~\ref{fig:fig18} illustrates a conceptual human-in-the-loop AIT architecture combining automated handling, human intervention, and functional verification.

Realizing such an architecture would require the satellite, AIT process, MGSE, EGSE, and automation equipment to be developed concurrently from the early design stage, consistent with concurrent-engineering approaches that integrate the development of space products and their associated life-cycle processes \cite{loureiro}. AIT automation should therefore be treated as a satellite-and-process co-design consideration rather than being introduced only after the satellite architecture and interfaces have been fixed.
\section{Conclusion}
This paper presented an integrated co-design framework for the assembly and integration of small satellites. The approach has been implemented on the 65~kg Lumelite-5 SM as an illustrative example. In this work, the satellite’s architecture, subsystem placement, structural interfaces, harness routing, and ground-support equipment were developed concurrently instead of independent activities. The resulting framework comprises centralized service interfaces, skeletal support interfaces, a removable double-hinge mechanical ground support equipment mechanism, and dedicated harness-routing and tie-down provisions.

The implemented features support accessibility throughout progressive satellite integration. The centralized service interfaces provide a common accessible region for EGSE connection and checkout, while the separate battery charging interface enables battery charging with the satellite assembled and stowed. The skeletal support interfaces provide direct access for component installation and harness routing during initial assembly while retaining access after integration without disturbing the established harness configuration. The double-hinge MGSE similarly provides access to bottom-deck subsystems and components during both initial assembly and subsequent integration activities, with the top deck supported at opening angles of 90$^\circ$ and 180$^\circ$. The dedicated harness-routing and tie-down provisions provide defined cable paths and restraint locations to maintain an organized and repeatable harness configuration throughout progressive integration. The use of an EGSE extension harness allows bus-level functionality to be retained in it’s configuration while the top deck is open to support functional checks, troubleshooting, burn-in, and staged build-and-test activities with the hardware physically accessible.

Together, the four co-designed techniques provide a structured approach to progressive assembly, intermediate functional verification, troubleshooting, rework, and final satellite closeout while minimizing unnecessary disturbance to previously integrated hardware. The Lumelite-5 SM implementation illustrates that AIT efficiency can be addressed as a satellite-architecture and process co-design consideration from the early development stage, rather than as a downstream activity after the satellite configuration has been fixed. This approach provides a basis for more accessible, repeatable, and scalable AIT workflows in future small-satellite and constellation programs.

\acknowledgements
The Lumelite-5 program is supported by the Agency for Science, Technology, and Research (A*STAR) under the RIE2020 Advanced Manufacturing and Engineering IAF-PP (project No. A1897a0040). The authors would also like to thank the past and present research staff and faculty members of NUS STAR and A*STAR who contributed, either directly or indirectly, to the Lumelite-5 programme.

The authors used ChatGPT and Claude for grammar checking and improving the clarity of Sections 2--5 in the manuscript.
\bibliographystyle{IEEEtran}

\thebiography
\begin{biographywithpic}
{Sen Hao Kam}{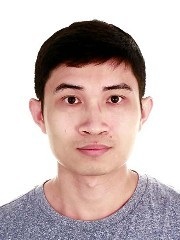}
\textit{received a B.Eng. degree in Mechanical Engineering from Nanyang Technological University (NTU), Singapore. He began his career as a Design Engineer, where he spent five years developing consumer products, including air purifiers and hair straighteners, with a focus on mechanical design, product development, and manufacturing. He subsequently joined the Satellite Technology and Research Centre (STAR) at the National University of Singapore (NUS) as a Senior Design Engineer. His work focuses on the mechanical design, structural analysis, assembly, integration and testing (AIT), supporting ground equipment. His research interests include design for manufacturability, satellite assembly and integration and mechanical ground support equipment.}
\end{biographywithpic} 

\begin{biographywithpic}
{Kay Soon Low}{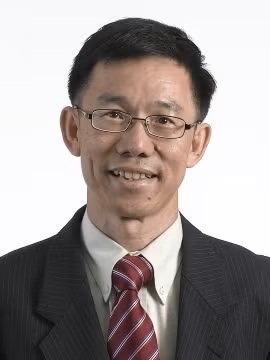} 
\textit{(Life Senior Member, IEEE) received the B.Eng. degree from the National University of Singapore, and the Ph.D. degree from the University of New South Wales, Sydney, Australia, both in Electrical Engineering. He has been a Professor with the Department of Electrical and Computer Engineering, National University of Singapore since 2016. He is also the founding director of Satellite Technology and Research centre (STAR). After completion of 4 satellites (Lumelite-1 to Lumelite-4), he transitioned to an advisory role at STAR in Dec 2023 while continuing his Professorship in the University. Throughout his career, Dr. Low has bridged academia and industry, consulting for numerous corporations and securing patents on nonlinear circuits, UWB systems, solar energy management system, imaging sensor and satellite system. Before his present appointment, he was the founding director of SaRC (Satellite Research Centre) with the School of Electrical and Electronic Engineering, Nanyang Technological University between 2009 and 2016. Under his leadership, his satellite team in NTU \& NUS has successfully completed 12 satellites. Dr. Low received Defence Technology Prize (Research \& Development category) in 2011, the IEEE Industrial Electronics and Applications (ICIEA) 2014 best conference paper award, the Singapore Public Administration Medal (Bronze) in 2014, and the joint team award for the President’s Technology Award in 2016. } 
\end{biographywithpic}

\begin{biographywithpic}
{Sheral Crescent Tissera}{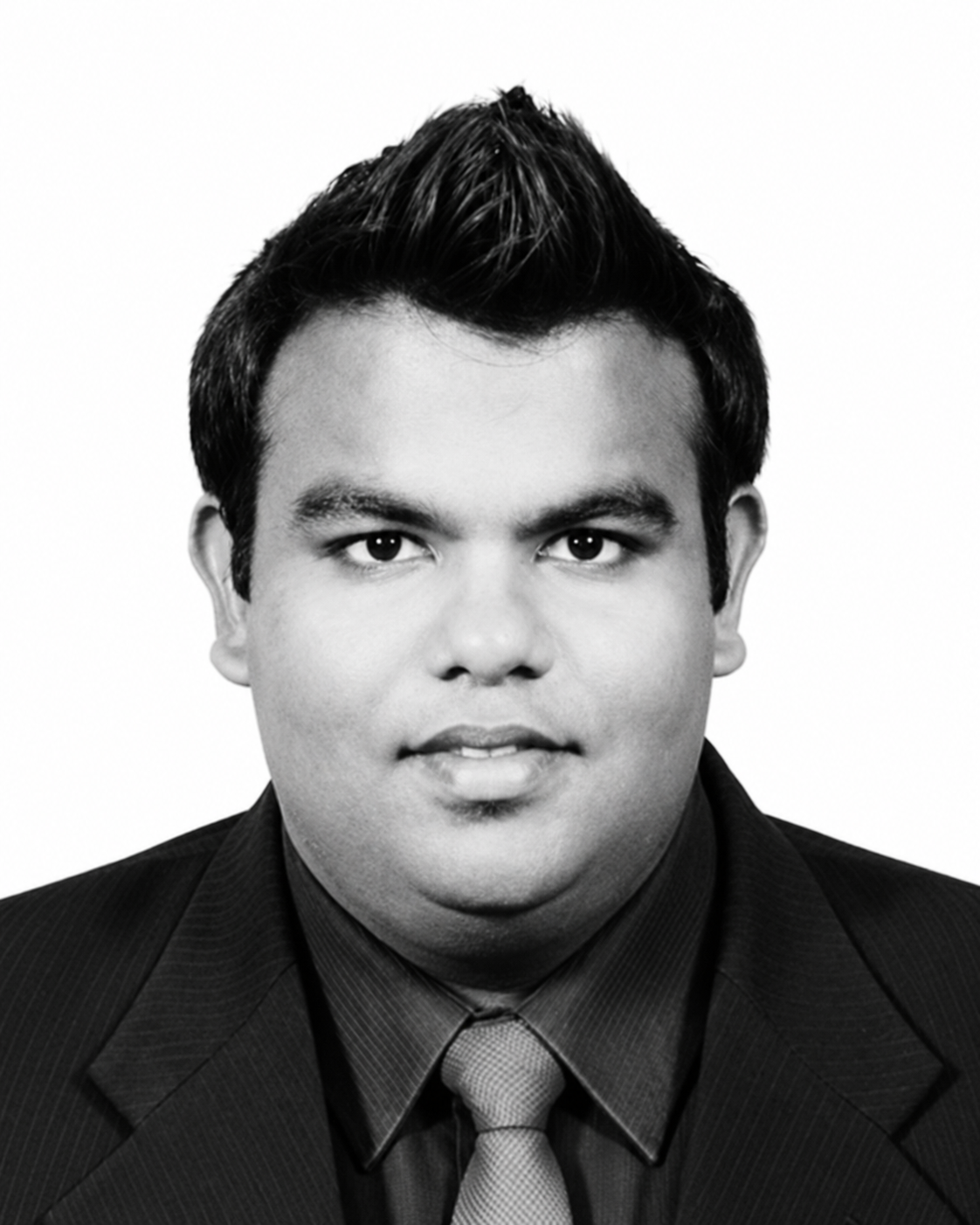}
\textit{is an Assistant Professor in the Department of Electrical and Computer Engineering, National University of Singapore (NUS), with a Courtesy Joint Appointment in the Department of Mechanical Engineering. He is a founding member of the Satellite Technology and Research Centre (STAR), a joint laboratory between NUS and DSO National Laboratories, and currently serves as its Deputy Director. Prior to joining NUS, he was a Research Engineer at the Satellite Research Centre (SaRC), Nanyang Technological University (NTU), Singapore, from 2013 to 2017. He received his B.Eng. in Aerospace Engineering and M.Eng. in Electrical and Electronic Engineering from NTU in 2013 and 2017, respectively, and earned his Ph.D. in Electrical and Computer Engineering from NUS in 2023, focusing on satellite attitude determination and control. His research centres on guidance, navigation, and control (GNC) for space autonomy, with particular emphasis on spacecraft attitude determination and control, space-based surveillance for space situational awareness, and collision avoidance. He also maintains complementary research interests in small satellite systems, space robotics, space mechanisms, edge computing, and GNC verification and experimentation through hardware-in-the-loop testing.}
\end{biographywithpic}

\end{document}